\documentclass[a4paper, titlepage,twocolumn]{article}

\usepackage{amsmath}
\usepackage{graphicx}
\usepackage{float}
\usepackage{multicol}
\usepackage{setspace}
\usepackage[round,colon]{natbib}

\floatstyle{ruled}
\newfloat{aBox}{thp}{lop}
\floatname{aBox}{Box}

\begin{document}

\title{Noise characteristics of the {\it Escherichia coli} rotary motor}
\author{Diana Clausznitzer$^{1,2,3}$ and Robert G. Endres$^{1,2, *}$}
\date{$^1$~Division of Molecular Biosciences, Imperial College London, SW7 2AZ London, UK; $^2$~Centre for Integrative Systems Biology at Imperial College, Imperial College London, UK; $^3$~BioQuant, Universit\"at Heidelberg, 69120 Heidelberg, Germany\\
$^*$~e-mail: r.endres@imperial.ac.uk\\
}

\maketitle

\begin{abstract}
The chemotaxis pathway in the bacterium {\it Escherichia coli} allows cells to detect changes in external ligand concentration (e.g. nutrients). The pathway regulates the flagellated rotary motors and hence the cells' swimming behaviour, steering them towards more favourable environments.
While the molecular components are well characterised, the motor behaviour measured by tethered cell experiments has been difficult to interpret.
Here, we study the effects of sensing and signalling noise on the motor behaviour. Specifically, we consider fluctuations stemming from ligand concentration, receptor switching between their signalling states, adaptation, modification of proteins by phosphorylation, and motor switching between its two rotational states.
We develop a model which includes all signalling steps in the pathway, and discuss a simplified version, which captures the essential features of the full model. We find that the noise characteristics of the motor contain signatures from all these processes, albeit with varying magnitudes. This allows us to address how cell-to-cell variation affects motor behaviour and the question of optimal pathway design. A similar comprehensive analysis can be applied to other two-component signalling pathways.
\\
{}  MWC model \slash{} two-component system
\end{abstract}

\section*{Introduction}
Biological systems sense stimuli from their environment using cell-surface receptors, and process this information to make reliable decisions, e.g. where to move, or whether to divide or to express new enzymes. Typically, intracellular signalling molecules are activated by modification, e.g. phosphorylation and methylation, and interact in complicated biochemical reaction networks.
Biochemical reactions rely on probabilistic collisions of a limited number of molecules. Hence, the number of signalling molecules fluctuates with time, making signal processing noisy. The abundance of noise sources in a cell is in stark contrast to the remarkable accuracy with which cells are known to respond to minute amounts of chemical concentration, including growing axons and immune cells~\citep{SykEisen96,MortGood10}.

The high biological relevance of noise has widely been recognised and studied extensively in gene expression~\citep{EloLevSwain02,PedOud05,Paul2005,Acarpanarnelovanoud10,EldElo10}. In contrast, noise in signal transduction is not well characterised, despite its importance for accurate sensing and cell-decision making.
Examples of eukaryotic systems, in which signalling noise has been considered include the ultrasensitive thresholding cascades~\citep{ThatOud02}, pheromone sensing in yeast~\citep{ColGorBrent05,TayFalHan09}, signal transduction in photoreceptors~\citep{DetRamShr00} and feedback loops for noise suppression~\citep{HornBar08,Lesvinpau10}. Furthermore, signalling noise has been considered in parts of bacterial pathways~\citep{EmoClu08,MethaGoyWin08,MoraWin10}. However, analyses have either been not comprehensive, or signal and noise transmission have not been compared explicitly.

An important class of signalling pathways are the bacterial two-component systems, including
\begin{figure}[!h]
  \centering
\includegraphics[width=0.8\columnwidth]{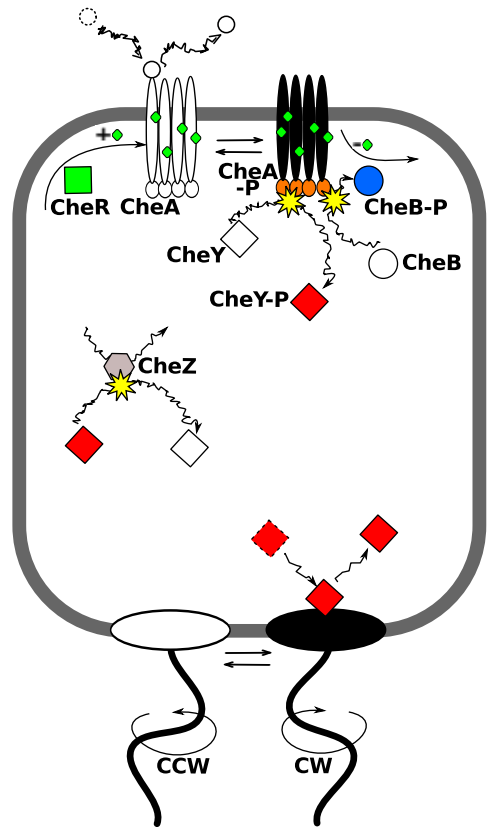}
     \caption{\label{fig:pathway}Chemotaxis pathways of {\it E. coli}. The pathway consists of transmembrane chemoreceptors, which constantly undergo molecular transitions between their {\it on} (black) and {\it off} (white) conformations. Receptors signal to CheA, which subsequently autophosphorylates. The average receptor activity is modulated by ligand binding and unbinding, as well receptors methylation and demethylation. CheA phosphorylates the response regulator CheY, which defuses through the cell and binds to the rotary motors. Upon CheY-P binding, motors switch from their default state of CCW rotation (i.e. running mode) to CW rotation (i.e. tumbling mode). In addition, CheY-P is dephosphorylated by its phosphatase CheZ. Receptor methylation is catalysed by CheR, which preferentially modifies inactive receptors. Receptor demethylation is catalysed by CheB, which is activated by phosphorylation, and modifies preferentially active receptors.}
\end{figure}
hundreds of pathways responsible for wide ranging functions such as sensing of and responding to nutrients, osmolarity, antibiotics, as well as quorum signals~\citep{LaubGou07}. A particularly well characterised example is the chemotaxis pathway in {\it E. coli}~(Fig.~\ref{fig:pathway}), allowing cells to swim towards nutrients and away from toxins with high sensitivity over a wide range of ambient concentrations~\citep{Berg00,FalHaz01,Sou04,WadArm04,BakWolStock06}. Specifically, the kinase CheA autophosphorylates when receptors are active and passes on phosphoryl groups to the response regulators CheY and CheB. Phosphorylated CheY~(CheY-P) modulates the probability of counterclockwise~(CCW) or clockwise~(CW) rotation of the motor. The rotational directions of motors correspond to the two swimming modes of the bacterium, namely smooth swimming and tumbling, respectively. Adaptation, i.e. the reversal of the effect of changes in the ligand concentration, is mediated by reversible receptor methylation and demethylation, catalysed by enzymes CheR and phosphorylated CheB~(CheB-P), respectively.

Using the {\it E. coli} chemotaxis pathway as an example, we are interested in the  behaviour of the rotary motor, i.e. the cell's final output, and how its rotation is affected by signalling and noise.
Specifically, there are several fundamental questions we would like to address:

Firstly, what types of signals are transmitted and what types are attenuated by the pathway? Early work showed that the system responds to the time-derivative of the input signal~\citep{BloSegBerg82}.
A number of research groups have measured the averaged response of cells to chemotactic signals~\citep{BloSegBerg82,SegBloBerg86,ShimTuBerg10}, and found that slowly, as well as rapidly changing input signals are not transmitted by the pathway. The response to slowly changing signals is attenuated by adaptation, which reverses the activation by ligand binding~\citep{BloSegBerg83,TuShimBerg08,ShimTuBerg10}. 
Rapidly changing signals were conjectured to be attenuated by a third-order filter~\citep{BloSegBerg82,SegBloBerg86}. While the phosphorylation dynamics of CheY-P has been shown to contribute a first-order filter~\citep{TuShimBerg08}, the exact filtering dynamics of the full pathway has not been addressed.

Secondly, how is noise generated, amplified or filtered in the signalling pathway, and how do different sources of noise affect the motor behaviour?
The power spectrum, which captures the correlations between fluctuations in motor behaviour at different time points, was measured for wild-type cells and mutant cells lacking the chemotaxis signalling pathway~\citep{KorEmClu04}. The spectrum was found to have a large low-frequency component in the wild-type cells, indicating that there is a dominant noise source in the signalling pathway with long correlations. \citet{KorEmClu04} and \citet{EmoClu08} showed, using simulations of the signalling pathway, that the adaptation dynamics plays an important role in generating long correlations. However, they only analysed the signalling pathway up to CheY-P.
Other studies include stochastic simulations of the noisy biochemical reactions of the pathway~\citep{MorFirthBray98}, and addressed the mechanism of motor rotation~\citep{XingBaiOster06,MeacciTu09, MoraYuWin09a, MoraYuWin09b,vanAlbtenWolde09}, including the thermodynamics of motor switching~\citep{ScharfFahBerg98,TurnSamBerg99,TuGrin05}. 
However, noise generation, filtering and amplification has not been addressed systematically for the various levels of the signalling pathway from chemoreceptors to motors.

Finally, how reliably are signals transmitted in the presence of noise?
An important task for the cell is to generate an appropriate response to input signals in the presence of fluctuations in the input, as well as due to noise in the biochemical signalling pathway.
Furthermore, cell-to-cell variation in protein expression influences signal transmission and noise filtering. Comparing these two aspects  of the pathway dynamics, the signal-to-noise ratio, is a novel perspective in our present study.

In the following, we present a mathematical model for the chemotaxis signalling pathway. A simplified pathway is given in the main text, while details of the full pathway are provided in the Supplementary Information ({\it SI}). We discuss the average~(deterministic) response of the signalling pathway to concentration signals. We analyse the noise sources in the signalling pathway and their effects. Finally, we vary pathway parameters and study how they affect signal and noise transmission. We also discuss briefly how our approach can be applied to other two-component systems and signalling pathways. Introductions to our modelling approach (Box~1), the mathematical characterisation of signal and noise propagation (Box~2), as well as a comparison of {\it E. coli}'s chemotaxis pathway and other two-component pathways (Box~3) are also given.

\section*{Results}

\subsection*{Experimental measurements of response and noise spectrum}

\begin{figure*}[t]
     \includegraphics[width=0.98\textwidth]{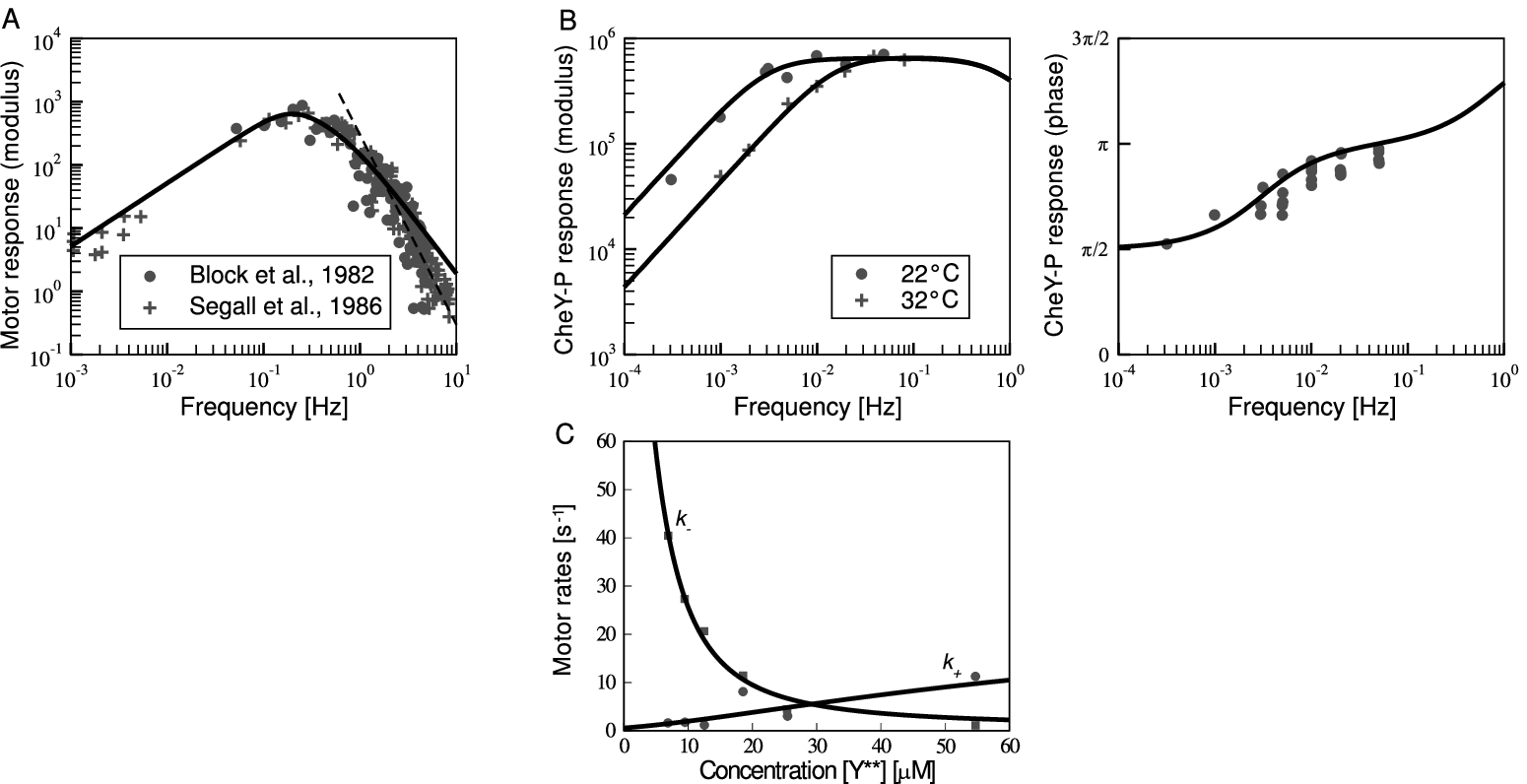}
     \caption{\label{fig:Shimizu}\label{fig:1982Block1986Segall} Calibration of the model. (A and B) Fourier transformed linear response function.~(A) Magnitude of the response function of the rotary motor measured by \citet{BloSegBerg82} (circles) and \citet{SegBloBerg86} (plus symbols). The fit of our model is shown by the solid line. The dashed line represents a 3rd-order filter for comparison.
     (B) Magnitude~({\it left}) and phase~({\it right}) of the response function at the level of the response regulator CheY measured by \citep{ShimTuBerg10}. The symbols are measurements at 22$^{\circ}$ (circles) and 32$^{\circ}$ (plus symbols). Grey lines represent the fit of our model to the magnitude of the response.
     (C) \label{fig:switchrates}Switching rates of the motor from CCW to CW rotation $k_+$ (squares) and from CW to CCW rotation $k_-$ (circles) as a function of the concentration of signalling-active mutant Y$^{**}$. A fit using the model of \citet{TurnSamBerg99} is shown as well (solid and dashed lines; cf. {\it SI}).
}
\end{figure*}

The signal propagation in the chemotaxis pathway has been characterised by the response to small concentration signals (linear response function; see Box~2). Specifically, the response has been measured at the level of \mbox{CheY-P} using fluorescence resonance energy transfer (FRET) by \citet{ShimTuBerg10}. In that study the system was stimulated by a periodic variation of the concentration of attractant MeAsp. Using a series of frequencies of the stimulation, the magnitude (modulus) and phase, i.e. the lag between signal and response, of the response was determined. In cell-tether experiments of motor rotation, the response to short impulses of attractants was measured at the level of the motor by \citet{BloSegBerg82} and \citet{SegBloBerg86}. Such data determines the linear response function up to a constant factor. Experimental results are shown in Fig.~\ref{fig:Shimizu}.
Noise propagation in wild-type cells has only been studied at the level of the motor by~\citet{KorEmClu04}.
We use the experimental response functions to calibrate our model, and subsequently study the noise power spectrum and signal-to-noise ratio.

\subsection*{Simplified model for the pathway}

Here we consider a simplified pathway to gain intuition of the key processes involved. 
The simplified pathway consists of chemoreceptor signalling in response to ligand binding and receptor methylation, as well as the rotary motor. Specifically, we use stochastic differential equations in a Langevin approach (\citealp{vanKampen2007}; see Box~1) to describe the dynamics of each type of signalling protein. We assume throughout that fluctuations in concentration are small, allowing us to describe the average behaviour of a signalling molecule by a deterministic dynamics and the fluctuations around the mean by additive noise.

\begin{aBox*}
\caption{Langevin approach for signalling in the pathway.}

\begin{multicols}{2}
\small
\setlength{\parindent}{1em}

In this paper, we use the Langevin approach to describe the noisy dynamics of signalling. This approach is based on the assumption that on average the dynamics of signalling follows a deterministic equation, an ordinary differential equation (ODE) for the rate of change. Consider for instance the following (linear) ODE for the rate of change of $R(t)$, which could describe the phosphorylated fraction of a molecular species,
\begin{equation*}
 \frac{d R}{ dt } = k - \tau^{-1} R.
\end{equation*}
This equation describes how $R$ changes due to constant, basal production with rate $k$ and decays with rate $\tau^{-1} R$, i.e. proportionally to the number of molecules $R$. The latter term results in the exponential decay of $R$ (similar to radioactive decay) with a characteristic time constant $\tau$ (or characteristic frequency $\tau^{-1}$).

The Langevin approach assumes that fluctuations around the average dynamics are small and well described by a noise term $\eta(t)$ added to the above (deterministic) dynamics. The resulting equation is a stochastic differential equation (SDE),
\begin{equation*}
 \frac{d R}{ dt } = k - \tau^{-1} R + \eta(t).
\end{equation*}
The noise term $\eta(t)$ is typically assumed to be Gaussian and white. The first property means that fluctuations around the average value of $R$ are assumed to be normally distributed. The second property means that the noise contains all frequency components or colours, and specifically that the autocorrelation function (see Box~2) is $\langle \eta(t) \eta(t')\rangle = Q \delta(t-t')$. Here, $Q$ is the intensity of the noise and $\delta(x)$ the Dirac delta function. The angular brackets indicate averaging over numerous time series. This autocorrelation function indicates that fluctuations in the rate at any two different time points are not correlated.

\paragraph{Linearisation of the equations.}
In order to characterise the response to input signals, as well as fluctuations in Box~2, we need to linearise and Fourier transform the dynamical equations. Here, we illustrate this procedure for our simple example.
Considering our ODE from above, we note that the variable $R(t)$ assumes a constant steady-state value $R^*$ when its rate of change is zero, i.e. $dR/dt = 0$. We obtain the dynamics of deviations of $R(t)$ from its steady state, e.g. due to an input signal, by linearising $R(t) = R^* + \Delta R(t)$ around the steady state. Inserting this linearisation into the above example yields
\begin{equation*}
 \frac{d [R^* + \Delta R(t) ]}{ dt } = k - \tau^{-1} [R^* + \Delta R(t) ].
\end{equation*}
As the rate of change of the steady state is zero by definition, we obtain for the dynamics of deviations $\Delta R(t)$
\begin{equation*}
 \frac{d [\Delta R(t)]}{ dt } = - \tau^{-1} \Delta R(t).
\end{equation*}

Considering the SDE, we are interested in the dynamics of fluctuations in $R(t)$, i.e. deviations from its average value $\langle R \rangle$. Hence, we write $R(t) = \langle R \rangle + \delta R(t)$, with $\delta R(t)$ fluctuations in $R$. Inserting into the SDE yields
\begin{equation*}
 \frac{d [\langle R \rangle  + \delta R(t) ]}{ dt } = k - \tau^{-1} [\langle R \rangle  + \delta R(t) ] + \eta(t).
\end{equation*}
The dynamics of the average $\langle R \rangle $ is described by the deterministic dynamics described by the ODE, whereas the dynamics of fluctuations is
\begin{equation*}
 \frac{d [\delta R(t)] }{ dt } = -\tau^{-1} \delta R(t) + \eta(t). 
\end{equation*}

\paragraph{Fourier transformed equations.}

Analysis of the dynamics is simplified by considering the Fourier transformed equations, defined for $\Delta R(t)$ as
\begin{equation*}
\Delta \hat{R}(\omega) = \int_{-\infty}^{\infty} \Delta R(t) e^{i \omega t} dt. 
\end{equation*}
Inserting the inverse Fourier transforms $\Delta R(t) = 1/(2\pi) \int_{-\infty}^{\infty} \Delta \hat{R}(\omega) e^{-i \omega t} d\omega $ into the ODE for $\Delta R(t)$ results in the equality
\begin{equation*}
-i \omega \Delta \hat{R}(\omega) =  -\tau^{-1} \Delta \hat{R} (\omega),
\end{equation*}
i.e. the time-dependent differential equation has been transformed into an algebraic equations, which can easily be solved for $\Delta \hat{R}(\omega)$. A similar equation can be obtained for the SDE for the fluctuations.

\paragraph{Determining the noise intensity.}
To discuss this aspect, consider the following general SDE:
\begin{equation*}
 \frac{dR}{dt} = r_{1} - r_{ 2} + \eta(t).
\end{equation*}
The rates $r_1$ and $r_2$ typically depend on the concentrations of proteins in the signalling network. The noise term $\eta(t)$ is composed of two terms $\eta_{1}(t)$ and $\eta_{2}(t)$, which are associated with the rates $r_{1}$ and $r_{2}$, respectively. We assume $\eta_{1}$ and $\eta_{2}$ to be independent, i.e. $\langle \eta_{1}(t)\eta_{2}(t') \rangle = 0$. In general, this is justified as different reactions are catalysed by different proteins.
Using $\langle \eta_{j} (t) \eta_{j} (t') \rangle = Q_{j} \delta(t-t')$, the noise intensities can be calculated if we make the assumption that fluctuations are due to so-called birth and death processes, i.e. creation and destruction of the molecules with average rates $r_1^*$ and $r_2^*$. Then the associated noise intensities are $Q_1 = r_1^*$ and $Q_2=r_2^*$~\citep{ThatOud02}. The intensity of the total noise $\eta(t)$ is the sum $Q=Q_1 + Q_2$ due to the independence of the two noises. As forward and backward rate are equal at steady state, $Q$ is twice the reaction rate in one direction at steady-state.

\end{multicols}
\end{aBox*}

We assume $N$ receptors form cooperative signalling complexes, which can switch between an active ({\it on}) and an inactive ({\it off}) state. Their activity $A$ is described by the Monod-Wyman-Changeux (MWC) model~\citep{SouBerg04,MelTu05,KeyEndSko06,EndWin06,EndSouWin2008, ClauOleEnd10}. The activity depends on the external ligand concentration $c$ at the receptor complex, as well as the methylation level $M$ of the complex as detailed in {\it Materials and Methods}.

We consider $N_C$ receptor complexes in a cell, and assume that each complex signals independently of the others. The total activity~$A_c$ of all receptors in a cell is determined by the sum over all signalling complexes~$j$.
The dynamics of the total activity is
\begin{equation}
\frac{dA_c}{dt} = \sum_{i=1}^{N_C} \frac{\partial A}{\partial M} \frac{dM_j}{dt} + \frac{\partial A}{\partial c} \frac{dc_j}{dt} + \eta_{A_j} (t),\label{eq:activity} 
\end{equation}
i.e. the dynamics of the complex activity is affected by changes in the receptor complex methylation level~(first term), changes in ligand concentration~(second term), as well as fluctuations due to the switching of the complex between its states~(last term).
All noise terms~$\eta(t)$ introduced in this section are discussed in {\it Materials and Methods}.

Changes in the concentration originate from time-varying input signals $\langle c(t)\rangle$, as well as fluctuations due to ligand diffusion. The dynamics of the concentration at the $j$th receptor complex is given by
\begin{equation}
\frac{dc_j}{dt} = \frac{d \langle c(t) \rangle}{dt} + \eta_{c_j}(t),\label{eq:ligand}\\
\end{equation}
where the first term captures average concentration changes (indicated by angular brackets $\langle \cdots \rangle$), affecting all receptors, and the second term describes concentration fluctuations at each receptor complex.

Adaptation is provided by reversible receptor methylation and demethylation, whose dynamics is described by the following equation~\citep{ClauOleEnd10}:
\begin{equation}
\frac{dM_j}{dt} 	= \gamma_R (N-A_j) - \gamma_B A_j^3 + \eta_{M_j} (t).\label{eq:methylation}\\
\end{equation}
The total methylation level $M_j$ of a receptor complex $j$ is changed by methylation of receptors in the inactive state~(first term) and demethylation~(second term). This latter rate is assumed to be strongly dependent on the receptor complex activity as only active receptors are demethylated by phosphorylated demethylation enzymes. The last term represents fluctuations due to the noisy processivity of the methylation and demethylation enzymes.  

The motor is described as a two-state system with CW and CCW rotating states, corresponding to running and tumbling modes, respectively. The dynamics of the probability of tumbling mode (tumble bias) $P_\text{CW}$ is described by
\begin{equation}
\frac{dP_\text{CW}}{dt} = k_+(A_c) (1-P_\text{CW}) - k_-(A_c) P_\text{CW} + \eta_{P_\text{CW}} (t),\label{eq:motor} 
\end{equation}
where the first term represents the switching from CCW to CW with the transition rate $k_+$, the second term represents switching from CW to CCW with transition rate $k_-$ and the third term describes temporal fluctuations in switching rates.
Here, transition rates are modulated by the receptor signalling activity $A_c$, whereas in the full pathway model CheY-P modulates motor switching. These rates have been experimentally measured using signalling mutants expressing varying amounts of constitutively active signalling molecule CheY~\citep{TurnSamBerg99}. The switching rates, including a fit of the model we used~(\citealp{TurnSamBerg99}; cf. {\it SI}) to the data, are shown in Fig.~\ref{fig:switchrates}.

\subsection*{Signal propagation\label{sec:response}}

We consider the response to concentration signals at various levels in the signalling pathway to study how signals are transmitted to the rotary motor (see Box~2 for an introduction to the formal characterisation of the response).
Briefly, an input signal $\Delta c(t)$ is a concentration change relative to a constant background concentration $c_0$, affecting all receptors equally and representing a ``meaningful'' input to the chemotaxis signalling pathway. Hence, the concentration is given by $\langle c(t) \rangle = c_0 + \Delta c(t)$.
Furthermore, cells are assumed to be adapted to the pre-stimulus concentration $c_0$ with the various levels~$R$ of the signalling pathway adapted to their steady-state values $R^*$.
\begin{aBox*}
\caption{Characterisation of signal and noise propagation.}
     \hspace*{2cm}\includegraphics[width=0.7\textwidth]{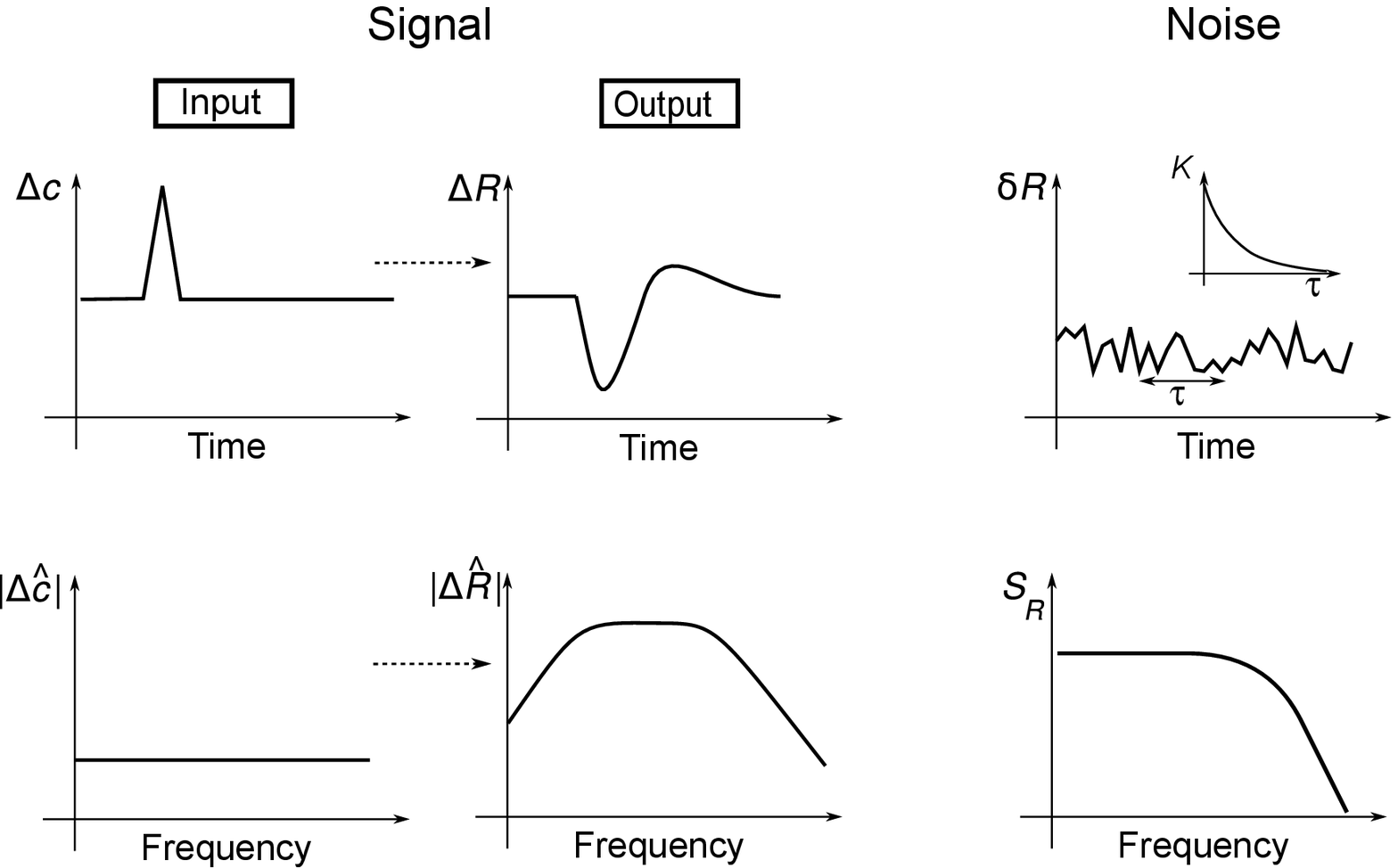}

\begin{multicols}{2}
\small
\setlength{\parindent}{1em}

\paragraph{Signal response.} The response of a system to an arbitrary small signal is described in terms of a characteristic function of the system, the linear response function $\chi_R$.
The time-dependent response $\Delta R(t)$, i.e. the deviation from the adapted state due to a small input signal $\Delta c(t)$, is linear and determined by
\begin{equation*}
 \Delta R(t) = \int_{-\infty}^t \chi_{R}(t-\tau) \Delta c(\tau) d\tau.
\end{equation*}
Hence, the time course of the response is determined by the convolution of the linear response function and the input signal. The linear response function describes the dynamics of the pathway and the convolution with the input signal represents the fact that the current state of the system is determined by the history of the input signal~\citep{Kubo57}. The Fourier transform of this equation reads more simply
\begin{equation*}
 \Delta \hat{R}(\omega)  = \hat{\chi}_R(\omega)  \Delta \hat{c}(\omega).
\end{equation*}
$\hat{\chi}_R(\omega)$ is also called the frequency-dependent gain~\citep{DetRamShr00}.
The magnitude $\vert\hat{\chi}_R(\omega)\vert$ describes what frequencies of the input signal are transmitted well, and which ones are attenuated.

Typically, finite activation rates of the system limit the response to rapidly changing input signals, i.e. high-frequency signals.
In this case, the Fourier transformed linear response functions falls off at high frequencies, and the system is called a low-pass filter.
If low-frequency components of the input signal are filtered out rather than high-frequencies, the system is called a high-pass filter. The chemotaxis pathway is a band-pass filter (see figure), filtering out low and high-frequency components.

To obtain a succinct measure for the signalling response due to an input concentration change $\Delta c(t)$, we define $\Delta R^2$ the integral over the response over frequency
\begin{equation*}
 \Delta R^2 = \int_{-\infty}^{\infty} d\omega \vert \hat{\chi}_{R}(\omega) \Delta \hat{c}(\omega) \vert^2.
\end{equation*}

\paragraph{Noise.}
Typically, any system is subject to various sources of noise, i.e. random fluctuations in the input, as well as from the internal signal processing. This is true in particular for biological systems, which rely on biochemical reactions and conformational changes of signalling molecules, which are intrinsically probabilistic.

Fluctuations $\delta R(t)$ around the mean value $\langle R(t)\rangle$ can be characterised by their correlations at different time points. The autocorrelation function $K$ is defined as
\begin{equation*}
 K(\tau) = \langle \delta R(t) \delta R(t+\tau) \rangle,
\end{equation*}
i.e. the average value the product of fluctuations at two time points. It only depends on the interval between time points if the dynamics of $R(t)$ is stationary, i.e. the mean value $\langle R(t)\rangle$ and variance $\langle\delta R^2(t)\rangle$ do not vary with time $t$. Averaging over different measurements of $R(t)$ is indicated by angular brackets. Typically, correlations decay with the interval length $\tau$ between time points. Often, the power spectrum is used to characterise fluctuations. According to the Wiener-Kinchin theorem the power spectrum is the Fourier transform of the autocorrelation function~\citep{Str67},
\begin{equation*}
 S_R(\omega) = \int_{-\infty}^{\infty} K(t) e^{i \omega t} dt.
\end{equation*}
For exponentially decaying correlations as in the figure, the power spectrum is Lorentzian, i.e. has the frequency dependency
\begin{equation*}
 S_R(\omega)\propto \frac{1}{\omega^2+\omega_\alpha^2}.
\end{equation*}
The power spectrum can be calculated from the absolute square of the Fourier transform of time series~$\delta R(t)$ measured or simulated over a time interval $T$,
\begin{equation*}
 S_R(\omega) = \lim\limits_{T\to\infty} \frac{\langle \delta\hat{R}(\omega) \delta \hat{R}^*(\omega) \rangle}{T},
\end{equation*}
where the Fourier transformation is defined on the finite measurement interval $T$ and the average $\langle \cdot \rangle$ is over multiple time series.

The variance of a stationary process can be calculated as the integral of the power spectrum over frequency,
\begin{equation*}
 \langle \delta R^2 \rangle = \frac{1}{2 \pi} \int_{-\infty}^{\infty} d\omega \, S_R(\omega).
\end{equation*} 

\end{multicols}\end{aBox*}

\paragraph{Analytical results for linear response functions.\label{sec:response:eq}}

We can analytically calculate the Fourier transformed linear response function from the dynamical equations Eq.~\eqref{eq:activity}-\eqref{eq:motor} without noise (see {\it Materials and Methods}). Knowing the response functions allows us to calculate the response to an arbitrary input signal (Box~2). Furthermore, we can analyse the filtering of the signal at each level of the pathway.
The Fourier transformed linear response function for the total activity of all receptors in a cell is
\begin{equation}
 \hat{\chi}_{A_c}(\omega) = \frac{ -i \omega  N_C \frac{\partial A}{\partial c} }{ \omega_M -i \omega}.\label{eq:chi_A}
\end{equation}
The receptor activity is a high-pass filter: The magnitude of the response function is small for frequencies $\omega$ below $\omega_M = (\gamma_R+3\gamma_B {A^*}^2)\partial A/\partial M$, which is the characteristic frequency due to adaptation. For frequencies above $\omega_M$ the response function is a constant, given by the number of receptor complexes $N_C$ participating in the response, and their sensitivity $\partial A/\partial c$ to ligand, evaluated at steady-state. The sensitivity is proportional to the receptor complex size $N$, i.e. it describes the amplification of the response of a single receptor.

Similarly, the Fourier transformed response of the motor is given by
\begin{equation}
 \hat{\chi}_{P_\text{CW}}(\omega) = \frac{ \omega_2 }{\omega_{P_\text{CW}} -i \omega} \hat{\chi}_{A_c} (\omega).\label{eq:chi_motor}
\end{equation}
The motor is a low-pass filter, i.e. its dynamics introduces a frequency-dependent response, which is constant below the characteristic frequency $\omega_{P_\text{CW}} = k_+^* + k_-^*$ of the motor due to the steady-state switching rates $k_+^*$ and $k_-^*$.  The parameter $\omega_2$ describes the sensitivity of motor switching with respect to changes in receptor activity ({\it Materials and Methods}). At frequencies above $\omega_{P_\text{CW}}$ the response is reduced.
From Eq.~\eqref{eq:chi_motor} it is obvious that receptors and motor are in a cascade: The motor response introduces a new filter proportional to $(\omega_i -i\omega)^{-1}$ which simply multiplies the response function of the response of the receptor activity.
The response functions of the full pathway including the phosphorylation reactions are shown in the {\it SI}.

For further analysis, we can write the Fourier transformed linear response function as
\begin{equation}
\hat{\chi}_R (\omega) = \vert \hat{\chi}_R (\omega) \vert e^{i \phi_R (\omega)},
\end{equation}
where $\vert \hat{\chi}_R \vert$ is the magnitude and $\phi_R$ is the phase of the response function, which characterise the amplitude and lag of the response behind the input signal, respectively.

\paragraph{Model calibration.} 
Figure~\ref{fig:Shimizu} shows experimental data for the response function, as well as the fits of our full pathway model.
For the fit of our model to the data by \citet{ShimTuBerg10}, we adjusted only the adaptation rates, as measurements were restricted to low frequencies.
The fit at 32$^{\circ}$~C yields the same adaptation parameters as obtained from fitting dose-response curves of adapting cells~\citep{ClauOleEnd10} (Fig.~\ref{fig:Shimizu}, {\it left}). The adaptation rates for room temperature are one order of magnitude smaller. Importantly, fitting to the magnitude of the Fourier transformed response yields a good fit for the phase of the response as well~(Fig.~\ref{fig:Shimizu}, {\it right}).

\citet{BloSegBerg82} and \citet{SegBloBerg86} measured the response of the motor using impulses of attractant. For our fit we adjusted adaptation and motor switching rates. Compared to the data by \citet{ShimTuBerg10} at the same temperature, adaptation rates are one order of magnitude larger, i.e. adaptation is faster in these experiments. The parameter $\omega_{P_\text{CW}}$ of the motor switching is 2.1/s, consistent with switching rates of about 1~Hz~\citep{BloSegBerg83}. It is not clear from where the difference in adaptation rates between the two sets of experiments originates. Besides different experimental conditions, it may be due to \citet{ShimTuBerg10} using populations of cells, whereas measurements by \citet{SegBloBerg86} were done on single cells.
Fitted parameters are given in the {\it SI}.

\begin{figure*}[t]
  \centering
	\includegraphics[width=0.8\textwidth]{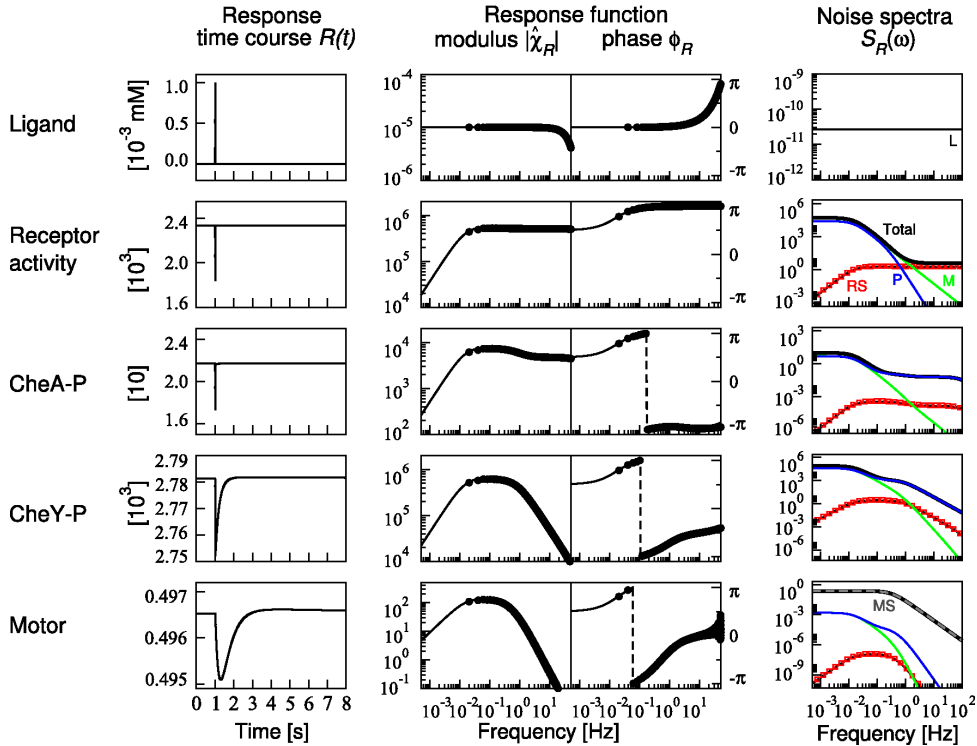}
     \caption{\label{fig:responsefunction}\label{fig:noise_spectra}Chemotaxis pathway response and noise spectra. ({\it Left} and {\it Middle}) Response upon impulse stimulation with attractant MeAsp. ({\it Left}) Time courses for MeAsp concentration $c$, total activity of receptors~$A_c$, CheA-P and CheY-P, and motor bias~$P_\text{CW}$. ({\it Middle}) Fourier transformed ligand signal, as well as response functions. Symbols correspond to numerical simulations and solid lines to analytically calculated response functions. ({\it Right})~Noise spectra of ligand and for the total activity of receptors~$A_c$, CheA-P and CheY-P, and motor bias~$P_\text{CW}$ (thick black lines). Also shown are the contributions to the spectrum from ligand binding~(L; thin solid black lines), receptor switching~(RS; thin red lines and symbols), receptor methylation and demethylation (M; green lines), as well as phosphorylation and dephosphorylation of CheA, CheY and CheB~(P; blue lines) and motor switching~(MS; dashed grey line).}
\end{figure*}

\paragraph{Signal filtering along the pathway.} 
Figure~\ref{fig:responsefunction} shows simulated time courses of the chemotactic response to an concentration impulse and the Fourier transforms of corresponding linear response functions.
We observe how the input signal is transmitted through the pathway, with the effective pulse durations becoming progressively longer along the pathway~(Fig.~\ref{fig:responsefunction}, {\it left}), including total receptor activity in a cell ($A_c$), phosphorylated kinase CheA, phosphorylated response regulator CheY, and finally the motor, characterised by its probability of tumbling ($P_\text{CW}$). In Fig.~\ref{fig:responsefunction} ({\it middle}) we show the corresponding linear response functions. 

The receptor activity acts as a high-pass filter, i.e. it transmits high-frequency signals, but not low-frequency signals. As can be seen from our simple model (cf.~Eq.~\eqref{eq:chi_A}), this property is due to adaptation, which introduces the time-derivative of the signal $\Delta c(t)$ up to the characteristic frequency $\omega_M$, eliminating the response to slowly changing attractant concentrations.
The activity of chemoreceptors is the input to further levels in the pathway.
The response of CheA-P is fast, and shows no qualitative difference to the response of receptors in the frequency range shown.
In contrast, due to the fast but finite rates of phosphorylation and dephosphorylation, preventing the CheY-P concentration to respond to rapidly changing input signals, the response at the level of CheY is reduced at high frequencies.
Similarly, the motor introduces another high-frequency filter due to slow switching between its two states. This additional filter can be deduced from Eq.~\eqref{eq:chi_motor}, where the motor response function takes the response of chemoreceptors as input, and additionally introduces a characteristic cut-off frequency $\omega_{P_\text{CW}}$ due to slow motor switching rates. 
Hence, the chemotaxis pathway acts as a band-pass filter~\citep{BloSegBerg82}, which only transmits input signals within a selected frequency range, which is of the order of 1 to 10~s. This time scale corresponds to the average time between two tumbles, allowing sensing of concentration changes during periods of running.

As shown in Fig.~\ref{fig:responsefunction}, {\it middle} the phase tends towards $\pi/2$, i.e. a quarter period, at low frequencies. This has been analysed by \citet{ShimTuBerg10} only for the receptor complex activity. This phase difference is due to adaptation and represents the fact that the system takes the time derivative of the stimulus below the characteristic frequency $\omega_M$ of adaptation.
The phase shift of the receptor activity increases to $\pi$ at high frequencies, indicating that the activity simply follows the output (a negative sign is due to the negative response of the activity to attractant concentration;~\citealp{ShimTuBerg10}).
The phase at high frequencies for the response of CheA follows the phase of the receptor activity, except for a small increase of the phase shift. In contrast, the phase of CheY and the motor increase significantly beyond $\pi$ indicating that slow rates of modification and motor switching introduce a lag of the response behind the stimulus.

\subsection*{Noise propagation\label{sec:noise}}
To understand the noise characteristics of the motor, we consider the noise sources and their transmission in the pathway. Each step in the signalling pathway is essentially probabilistic, hence, noisy: ligand diffusion and binding, receptor switching between its functional {\it on} and {\it off} states, as well as receptor methylation and demethylation, phosphorylation and dephosphorylation of signalling proteins CheA, CheY and CheB, and switching of the rotary motor between its two states, CW and CCW rotation. 
To characterise fluctuations of the phosphorylated signalling protein $\delta R(t)$ around its mean value $\langle R(t) \rangle$, we use the power spectrum $S_R(\omega)$ and the variance $\langle \delta R^2 \rangle = \langle R^2(t) \rangle - \langle R(t) \rangle^2$~(cf. Box~2).

\paragraph{Analytical results for noise spectra.}
Considering Eq.~\eqref{eq:activity}-\eqref{eq:motor} with noise, we can analytically calculate power spectra (see {\it Materials and Methods}). The power spectrum of activity fluctuations is given by
\begin{equation}
  S_{{A}_c}(\omega) = N_C \frac{ \omega^2 \left[ S_{a}(\omega) + \left(\frac{\partial A}{\partial c} \right)^2 S_{c}(\omega)\right] + \left( \frac{\partial A}{\partial M} \right)^2 Q_M}{\omega_M^2 +\omega^2}.\label{eq:Aspec_simple}
\end{equation}
In this equation we considered fluctuations from receptor switching (first term in numerator), ligand diffusion (second term), as well as the receptor methylation dynamics (third term) at each of the $N_C$ receptor complexes per cell. We have assumed that fluctuations at different receptor complexes are independent. Therefore, we obtain the sum of $N_C$ identical spectra for all complexes. The individual terms $S_{a}(\omega)$, $S_{c}(\omega)$ and $Q_M$ are given by Eq.~\eqref{eq:switchspec}, \eqref{eq:Cspec} and \eqref{eq:etaMspec} in {\it Materials and Methods}.
The frequency dependence of the ligand noise, as well as noise from receptor complex switching, indicates filtering of slowly varying fluctuations with frequencies below the characteristic frequency $\omega_M$ due to adaptation. In contrast, only high-frequency fluctuations from the receptor methylation dynamics are filtered by the adaptation dynamics. This is due to finite rates of methylation and demethylation fluctuations introducing correlations in the receptor methylation level.

The power spectrum of fluctuations in motor bias is obtained as
\begin{equation}
  S_{P_\text{CW}}(\omega) = \frac{ \omega_2^2 S_{A_c}(\omega) + Q_{{P_\text{CW}}}}{\omega^2 + \omega_{P_\text{CW}}^2} .\label{eq:Pspec_simple}
\end{equation}
The first term represents transmitted noise from receptor complexes, including the noise power spectrum of the receptor activity and the sensitivity of motor switching rates to changes in activity $\omega_2^2$. The second term is motor switching noise. Both noises are filtered by the motor, as its finite rates of switching introduce correlations with characteristic frequency $\omega_{P_\text{CW}}$.
The noise spectra of the full pathway including the phosphorylation reactions are shown in the {\it SI}.

\paragraph{Noise filtering along the pathway.}
In Fig.~\ref{fig:noise_spectra} ({\it right}), we show the power spectrum of fluctuations at the various levels of the signalling pathway, i.e. total receptor activity, CheA-P, CheY-P and the motor. We also plot the individual contributions from processes generating noise, namely ligand diffusion, receptor switching, methylation and demethylation of receptors, and phosphorylation and dephosphorylation of proteins, as well as motor switching. This allows us to follow how noise is generated and transmitted at the various levels of the pathway.
The noise spectrum of the receptor activity has its largest contribution at low frequencies, which originates in the receptor methylation and phosphorylation dynamics. Most of the fluctuations from phosphorylation stem from CheB (the separate contributions to the phosphorylation noise are not shown in Fig.~\ref{fig:noise_spectra}, {\it right}).
At high-frequencies, the activity noise spectrum is flat. This is due to ligand and receptor switching noise, which is removed at low frequencies by adaptation, but not at high-frequencies. The general behaviour of the noise spectrum corresponds to the simplified model~(cf. Eq.~\eqref{eq:Aspec_simple}).

The noise spectrum of CheA-P has generally the same shape as the activity spectrum with a large low-frequency component, mainly due to receptor methylation and CheB phosphorylation dynamics. This spectrum also has an almost flat high-frequency behaviour in the frequency range shown. Apart from ligand and receptor switching noise, the flat part of the spectrum is largely determined by fluctuations from CheA autophosphorylation, which has roughly the same shape as activity noise at high frequencies because autophosphorylation depends on the receptor activity.

The noise spectrum of CheY-P is also largest at low frequencies. However, at high frequencies the spectrum falls off as noise is filtered due to the finite rates of CheY phosphorylation and dephosphorylation, which introduce correlations in the fluctuations.

The motor introduces another layer of filtering of transmitted noise with the characteristic motor switching frequency~$\omega_{P_\text{CW}}$ (cf. Eq.~\eqref{eq:Pspec_simple}). Hence, transmitted noise is reduced by two filters in the frequency range shown, namely due to the CheY-P and motor dynamics. However, the main contribution to the spectrum is due to the motor switching itself, which is reduced only by a first-order filter with characteristic frequency~$\omega_{P_\text{CW}}$.

\section*{Cell-to-cell variation of motor behaviour\label{sec:vary}}

How are the signal response, fluctuations and the signal-to-noise ratio (SNR) affected by changing parameters of the pathway such as size of receptor complexes, protein concentrations and reaction rate constants?
In this section, we discuss the effect of cell-to-cell variation on the power spectrum of the motor. In the next section, we discuss the SNR and its contributions, and how they depend on receptor complex size and adaptation rates.

\begin{figure}[!h]
  \centering
     \includegraphics[width=0.95\columnwidth]{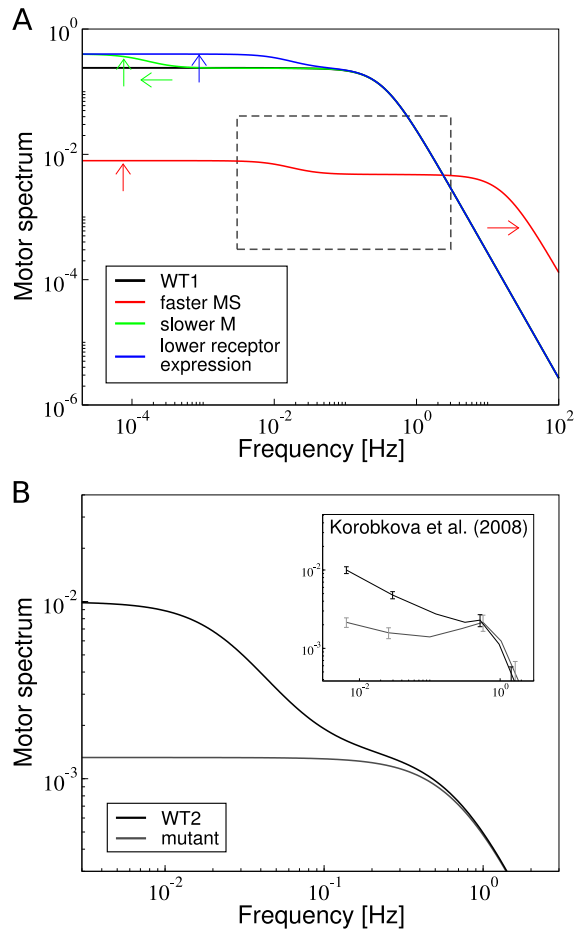}
     \caption{Effect of cell-to-cell variation on power spectrum of the motor. (A) Predictions about how different cell parameters affect the motor power spectrum, in particular its low-frequency component. The black line (wild-type WT1) is the same as the total motor spectrum in Fig.~\ref{fig:noise_spectra} ({\it right}). The motor spectra for increased motor switching rates (MS; red line), as well as reduced rates of receptor methylation and demethylation (M; green line) and the total number of receptors (blue line) are shown as well. Arrows indicate the features in the spectra that are affected. Dashed box is the area shown in panel (B).
     (B) Motor spectrum of cells with low motor bias (black line), as well as the component from steady-state motor switching only (grey line).
     ({\it Inset}) Measured power spectra for wild-type cells (WT2) with low motor bias (black) and mutant lacking the signalling pathways (grey). Axes are the same as in the main panel. Error bars indicate the measurement uncertainty. Spectra were traced from data presented by \citet{KorEmClu04}. Model parameters are listed in the {\it SI}.\label{fig:motorspec_prediction}}
\end{figure}

According to our model parameters obtained from fitting the Fourier transformed linear response to data, the main contribution to the power spectrum comes from the steady-state switching of the motor between CCW and CW state. However, cell-to-cell variation in protein content and motor switching rates can lead to modifications of the largely Lorentzian-shaped spectrum. These modifications are caused by the transmitted noise from receptor methylation and phosphorylation dynamics (green and blue lines in Fig.~\ref{fig:noise_spectra}, {\it right}). Specifically, Fig. ~\ref{fig:motorspec_prediction}A shows the motor power spectrum for increased motor switching rates as well as reduced adaptation rates and number of chemoreceptors in a cell. In all cases the low-frequency component of the transmitted noise becomes more prominent.

An increased low-frequency component has been observed in the motor power spectrum for cells with low motor bias by \citet{KorEmClu04}. These authors measured the motor power spectrum for wild-type cells, as well as mutants lacking the signalling pathway. The mutant's spectrum represents the component to the power spectrum from steady-state motor switching only. Wild-type cells showed a large low-frequency component compared to the mutants. Figure~\ref{fig:motorspec_prediction}B shows that our model can reproduce these experimental data (shown in the {\it Inset}).

\section*{Signal-to-noise ratio at the motor}

To characterise how signals are transmitted in the presence of noise, we define the SNR at the level of the motor as
\begin{equation}
 \text{SNR} = \frac{\Delta P_\text{CW}^2}{\langle \delta P_\text{CW}^2 \rangle}
\end{equation}
with $ \Delta P_\text{CW}^2$ and $\langle \delta {P_\text{CW}}^2 \rangle$ defined in Box~2. For optimal signalling this ratio should be maximised.
For simplicity, we only discuss the receptor activity in the text, while in the figures we additionally show the contribution from phosphorylation processes as transmitted to the motor.

\subsection*{Optimal receptor complex size}

\begin{figure*}[htp]
  \centering
     \includegraphics[width=0.8\textwidth]{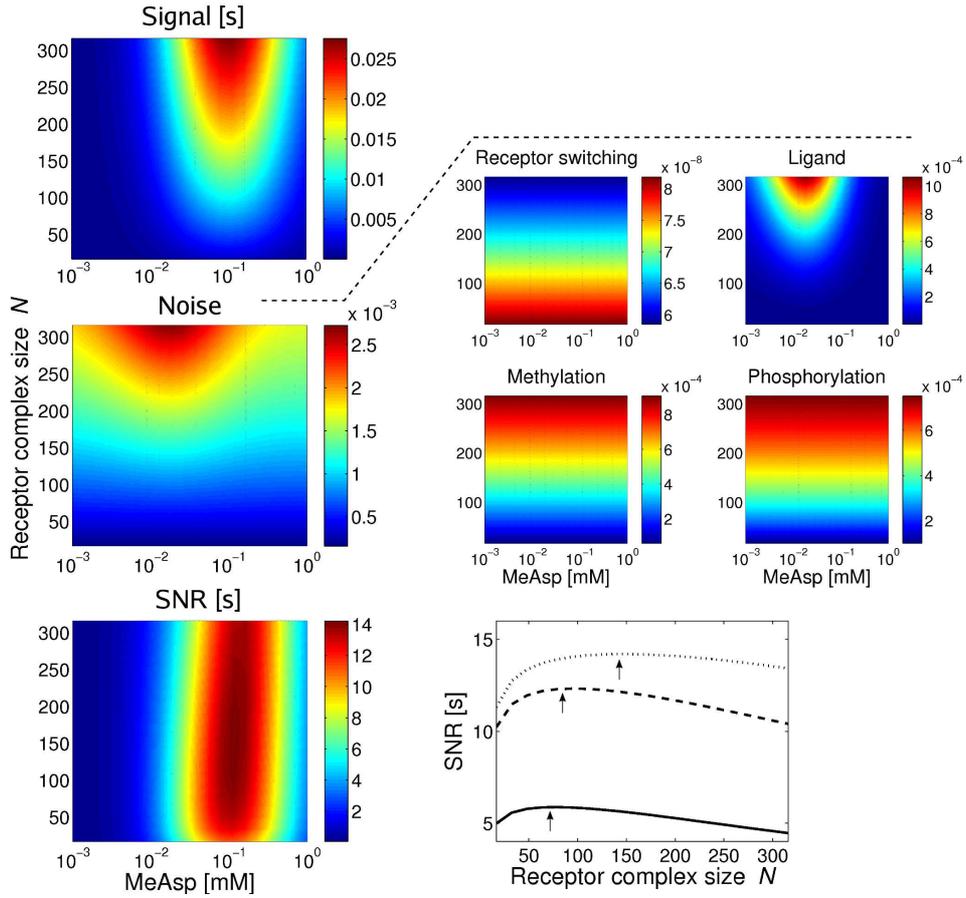}
     \caption[Varying ambient concentration and receptor complex size.]{Varying ambient concentration and receptor complex size.
({\it Top}) Integrated response of the motor bias.
({\it Middle}) Variance of the motor bias including only contributions from receptor switching, ligand diffusion, methylation and phosphorylation. The individual contributions are shown to the right of the main panel.
({\it Bottom, left}) SNR based on the signal response and variance shown in the top and middle panel.
({\it Bottom, right}) SNR as a function of receptor complex size at ambient concentration 0.02 (solid), 0.03 (dashed) and 0.05~mM (dotted line). Optimal complex size is indicated by arrows.
}\label{fig:SNR_vary_N}
\end{figure*}

Receptor complexes amplify small signals proportionally to their size $N$. However, also concentration fluctuations are expected to be amplified. Hence, we hypothesize that the receptor complex size could be optimised to yield a balance of advantageous amplification of signals and detrimental amplification of input noise.

In Fig.~\ref{fig:SNR_vary_N} ({\it top}) we show the integrated motor response $\Delta P_\text{CW}^2$ (see Box~2) to a step stimulus for varying background concentration and receptor complex size. We assume that the step stimulus size is a constant fraction of 10 percent of the background concentration.
The integrated response has a characteristic variation with background concentration with the maximum in the sensitivity range of Tar receptors (indicated by their dissociation constants). Furthermore, the response increases with receptor complex size~$N$.
We calculated the integrated signal response of the receptor activity (see {\it Materials and Methods}). This quantity scales linearly with receptor complex size, $\Delta A_c^2\propto N$, due to coherent addition of the signalling responses of different receptor complexes, amplification of concentration changes by receptor complexes, as well as filtering by adaptation.

In the middle panel of Fig.~\ref{fig:SNR_vary_N}, we show the variance (i.e., the integrated noise power spectrum, see Box~2) of the transmitted noise of the pathway at the level of the motor.
Only the contribution to the variance from ligand diffusion depends on the background concentration. Compared to the signal response, the maximum of the variance is shifted to a slightly lower concentration.
The contribution to the variance from switching of receptor complexes is relatively small compared to the other contributions and roughly constant with receptor complex size, whereas those from ligand diffusion, receptor methylation and phosphorylation dynamics increase with receptor complex size.

To understand these behaviours of the variance more intuitively, we analysed the receptor activity analytically (for details of the calculation, see {\it Materials and Methods}).
We find, the contribution to the variance of the receptor activity from receptor switching is indeed constant, independent of $N$. The contribution from ligand diffusion scales steeply as $N^2$, the difference between ligand noise and ligand signal amplification being due to ({\it i}) noise from different complexes is added up incoherently, and ({\it ii}) the main contribution to the variance coming from high-frequency ligand noise, which is not filtered by adaptation. The contribution from receptor methylation grows approximately linearly with receptor complex size as a result of the incoherent addition of fluctuations at different receptor complexes and the sensitivity of the receptor complex activity with respect to changes in methylation level increasing proportionally with $N$.
The contribution to the variance from phosphorylation processes grows with receptor complex size similar to the contribution from the methylation dynamics. 
Overall, the total variance of transmitted noise at the level of the motor has contributions from receptor switching, the dynamics of receptor methylation, and phosphorylation. The latter is approximately constant or grows slower than the amplified signal response, whereas the component from ligand diffusion increases steeper than the signal response with growing receptor complex size.

The resulting SNR, i.e. the ratio of integrated signal response and variance of the noise, is shown in Fig.~\ref{fig:SNR_vary_N} ({\it bottom}).
The SNR is largest at background concentrations in the sensitivity range of the Tar receptor. Furthermore, due to the different dependencies of the signal and the noise on the receptor complex size, the SNR has a maximum at a particular receptor complex size (Fig.~\ref{fig:SNR_vary_N} {\it bottom, right}). The SNR grows below that complex size due to signal amplification, while the amplified ligand noise from ligand diffusion is still below the internal noise level from receptor switching and receptor methylation and phosphorylation dynamics. Above the optimal receptor complex size, the SNR decreases because the ligand noise is amplified more than the signal.
\begin{figure*}[htp]
  \centering
     \includegraphics[height=0.8\textwidth]{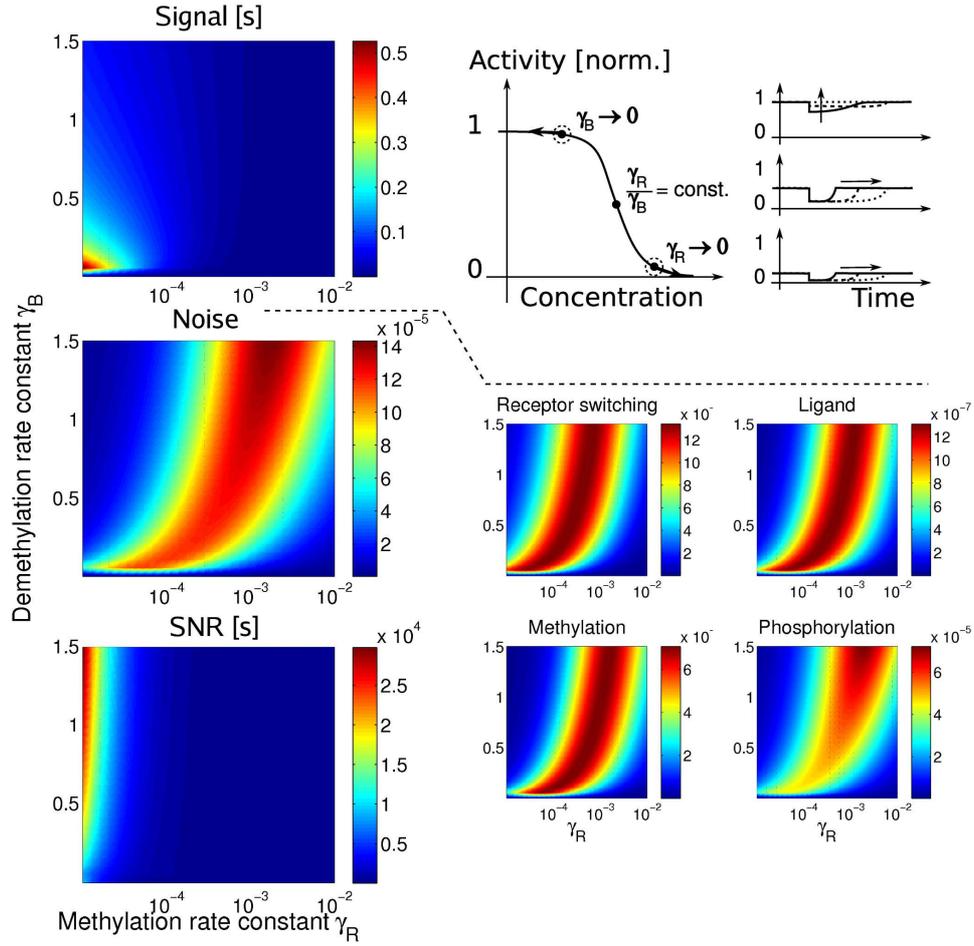}
     \caption[Varying receptor methylation and demethylation rate constants.]{Varying receptor methylation and demethylation rate constants $\gamma_R$ and $\gamma_B$, respectively.
({\it Top, left}) Integrated response of the motor bias. ({\it Top, right}) Illustration of the effects of vanishing $\gamma_R$ and $\gamma_B$ on adapted activity (indicated by dot and circle along dose-response curve; left), as well as on time courses (right) for three cases, $\gamma_R\to 0$ ({top right}), $\gamma_R/\gamma_B = \text{const}$ ({middle right}) and $\gamma_B\to 0$ ({bottom right}). For further explanation see text. 
({\it Middle, left}) Variance of the motor bias including only from receptor switching, ligand diffusion, methylation and phosphorylation. The four individual contributions are shown to the right of the main panel.
({\it Bottom, left}) SNR based on the signal response and variance shown in the top and middle panel. 
}\label{fig:SNR_vary_RB}
\end{figure*}

\subsection*{Optimal adaptation rates}
Adaptation filters slow input signals, its speed determining what input frequencies are transmitted by the pathway. Furthermore, the adaptation dynamics filters input noise. Hence, adaptation rates may be expected to be optimised for signal and noise propagation.

Figure~\ref{fig:SNR_vary_RB} shows the integrated signal response at the level of the motor for varying rates of receptor methylation ($\gamma_R$) and demethylation ($\gamma_B$). Varying these parameters describes changing the concentrations of receptor modification enzymes CheR and CheB. Interestingly, varying the two parameters has different effects on the signalling response: the integrated signal response increases for vanishing $\gamma_R$, whereas it decreases for vanishing $\gamma_B$.
There are two effects that contribute to this behaviour:
Firstly, if the concentration of one of the receptor modification enzymes is reduced, the receptors becomes modified predominantly by the opposing enzyme, hence driving the receptor activity towards saturation ($A^*=0$ or $A^*=1$). This effect would tend to quench the response by receptors. Secondly, as the enzyme concentration is reduced, adaptation times increase. Hence, this effect increases the integrated signal response as the time the receptor activity deviates from the adapted state increases.
According to calculations shown in {\it Materials and Methods} for the integrated response of receptors, the first effect dominates in the case of reduced $\gamma_B$: Due to the strong activity dependence of the demethylation rate, reducing the demethylation rate constant effects the adapted activity of receptors strongly. Hence, receptors are quickly driven into saturation for vanishing $\gamma_B$. In contrast, in the case of reduced $\gamma_R$ the second effect dominates and the increased adaptation time leads to an increased integrated signal response.
At large methylation and demethylation rates, adaptation times are reduced leading to a decreasing integrated signal response.

The variance of fluctuations is shown in the middle panel of Fig.~\ref{fig:SNR_vary_RB}. The individual contributions from transmitted noise at the level of the motor look qualitatively similar. All contributions decrease both for vanishing $\gamma_R$ and $\gamma_B$ consistent with calculations for the variance of the receptor activity in {\it Materials and Methods}.

The SNR is shown in the bottom panel of Fig.~\ref{fig:SNR_vary_RB}. The SNR increases for vanishing $\gamma_R$ and decreases for vanishing $\gamma_B$. According to Fig.~\ref{fig:SNR_vary_RB}, a large SNR is obtained for small $\gamma_R$ and large $\gamma_B$, corresponding to the parameters of our model.

\section*{Fluctuation-response relationships}

\citet{ParkPonClu10} presented the idea that the signalling response to concentration signals and fluctuations in the chemotaxis pathway are not independent of each other, because they are produced by the same molecular interactions. Specifically, based on measurements at the level of the motor these authors proposed a fluctuation-response theorem, namely an approximate linear relationship between the adaptation time to step stimuli and the variance of fluctuations in CheY-P concentration.

\begin{figure}[!h]
  \centering
     \includegraphics[width=0.95\columnwidth]{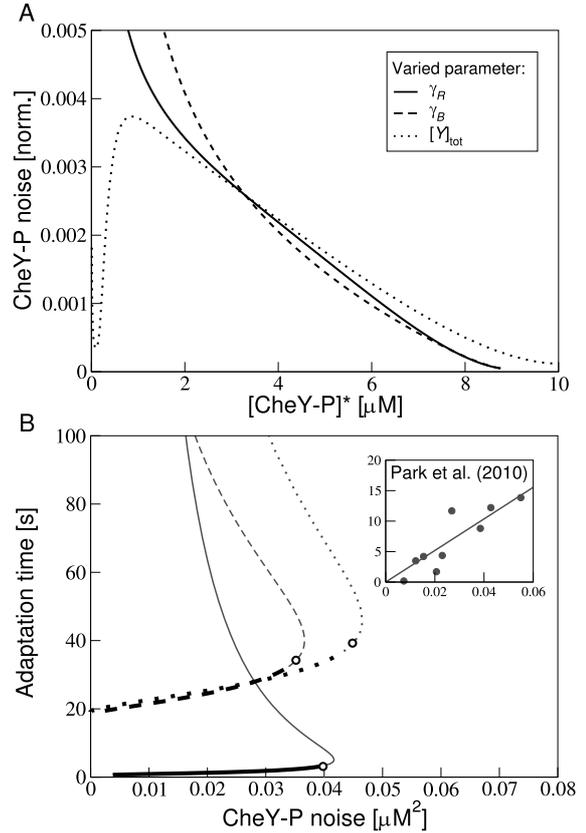}
     \caption{Fluctuation-response relationships. (A) Variance of CheY-P fluctuations (normalised by the squared adapted CheY-P value) as a function of adapted CheY-P value for varying adaptation rates $\gamma_R$ (solid line) and $\gamma_B$ (dashed line) and total CheY concentration (dotted line).
     (B) Adaptation time, calculated as $(\lambda_1 \partial A / \partial M)^{-1}$, plotted against the variance of CheY-P fluctuations. Adapted CheY-P concentration of 5 $\mu M$ is indicated by the circles. Legend is the same as in (A). Our parameters are given in the {\it SI}.
     ({\it Inset}) Variance of CheY-P concentrations plotted against the adaptation time as extracted from measurements of motor rotation by \citet{ParkPonClu10}. The line is a linear fit through the data.
     \label{fig:FRT}
     }
\end{figure}

Using our model, we tested this hypothesis and varied the adaptation rates, as well as the total CheY concentration in a cell, resulting in a shifted adapted CheY-P concentration at steady state. We find that the variance of CheY-P (normalised by the squared adapted value) decreases as the adapted CheY-P value increases except for very small adapted CheY-P concentrations (Fig.~\ref{fig:FRT}A), indicating that the relative strength of fluctuations decreases as expected.
In Fig.~\ref{fig:FRT}B we show the adaptation time, approximated by the inverse of the characteristic frequency due to adaptation, plotted against the variance of CheY-P. We find that at low adaptation times (thick line styles of the curves), the adaptation time increases with the variance of CheY-P, indicating that cells with large fluctuations also respond longer to concentration signals.
In contrast at long adaptation times, the adaptation time decreases with increasing variance of the pathway (grey parts of the curves). 
This behaviour can be directly traced back to the non-monotonic variance shown in Fig.~\ref{fig:SNR_vary_RB}.
It is maximal when the adapted CheY-P concentration is about 5 $\mu M$, i.e. when typically half of CheY is phosphorylated.
The exact relationship depends on what parameteris varied, exemplified by the different curves in Fig.~\ref{fig:FRT}B.
For each parameter and small adaptation times, we find an approximate linear relationship in line with \citet{ParkPonClu10}, see {\it Inset}.

\section*{Discussion}

Biological signalling pathways employ biochemical reaction networks and molecular state transitions to sense and process signals from the environment. Fluctuations inherent in these processes determine the signals which can reliably be transmitted. Here, we studied the signal and noise propagation in the {\it E. coli} chemotaxis signalling pathway, which controls the bacterial swimming behaviour in chemical gradients. Specifically, we considered the dynamics of ligand diffusion, receptor methylation and demethylation, receptor complex switching between {\it on} and {\it off}, phosphorylation and dephosphorylation of the kinase CheA, and response regulators CheY and CheB, as well as from rotary motor switching between CW and CCW direction. We assume cooperative chemoreceptor signalling complexes, whose activities depend on ligand concentrations and receptor methylation level, described by the MWC model~\citep{SouBerg04,MelTu05,KeyEndSko06,EndWin06,EndSouWin2008,ClauOleEnd10}.

We formulated a model which includes all processes in the signalling pathway, discussed in the {\it SI}. Not included is the dynamics of gene expression, which is assumed to be much slower than the dynamics of signalling processes. To make results intuitive we also presented a simplified version of the model, which only includes the dynamics of the activity of chemoreceptors, ligand concentration and receptor methylation level, as well as the motor dynamics.
To calibrate the model, we first collected experimental data sets for the signalling pathway and rotary motor~\citep{BloSegBerg82,SegBloBerg86,ShimTuBerg10}, and the motor switching behaviour~\citep{KorEmClu04,KorobEmClu06}. Using the Fourier transformed linear response function, we subsequently fitted our model parameters. We found a range of parameters fitting different data sets, revealing a striking experimental variation, which may require further characterization in the future.

Despite the fitting, there is a discrepancy of our response function and the data at large frequencies. \citet{BloSegBerg82} and \citet{SegBloBerg86} conjectured that the pathway is a third-order low-pass filter. In contrast, we find that the only relevant filters in that frequency range are due to CheY-P and motor dynamics, leading to only a second-order filter.
One explanation for the missing filter is that experimental concentration pulses were not short enough, leaving a signature from the input signal at large frequencies.
Alternatively, additional processes such as a slow release of CheY-P from the chemosensory complexes as discussed by \citet{BlatGillEis98} could lead to an additional filter.
However, CheY-P/CheZ complex formation and potential oligomerisation of CheY-P/CheZ complexes~\citep{BlatEis1996a,BlatEis1996b, EisenbachInEisenbach04} are not expected to contribute to high-frequency filtering ({\it SI}).

The motor behaviour is the final cell output, which contains characteristic noise signatures of all upstream signalling components, including the receptors. We found that motor switching is the dominant contribution to the spectrum of the fluctuations in motor bias. This may be not surprising as motor switching enables {\it E. coli} to tumble and change its swimming direction, and is therefore crucial for its search strategy.
However, low-frequency contributions from the dynamics of receptor methylation and phosphorylation processes may be dominant in particular cells~(Fig.~\ref{fig:motorspec_prediction}). We predict that due to cell-to-cell variation of protein contents or fast motor switching, these low-frequency components become important.  Specifically, \citet{KorobEmClu06} measured power spectra in cells with low motor bias and found an increased low-frequency component as compared to mutant lacking the signalling pathway. Our model is able to reproduce these spectra. Long correlations in motor bias may enable subpopulations of cells to swim further without tumbling or to tumble more frequently.

Although chemotaxis is one of many capabilities a cell has and may not be optimised in isolation without the rest of the cell, we speculate the cell aims to maximise the SNR for most efficient signalling and chemotaxis.
We found that the SNR is maximised at particular receptor complex sizes similar to values of receptor cooperativity extracted from FRET dose-response curves~\citep{EndSouWin2008}. In line with the data, the ``optimal'' complex size increases with external ligand concentration, and hence with receptor methylation level. While our complex sizes appear overestimated, noise from ligand molecules rebinding to the same receptor complex~\citep{BiaSet05} has not been considered here. This may well increase the noise level from external sources and hence decrease the predicted optimal receptor complex size.
Using our model, we also analysed the effect of varying the methylation and demethylation rate constants. 
We found that a large SNR is obtained for small methylation and large demethylation rate constant, corresponding to our fitted model parameters from FRET dose-response curves~\citep{ClauOleEnd10}.

\begin{aBox*}
\caption{Comparison of {\it E. coli} chemotaxis and other two-component systems.}

\begin{multicols}{2}

\includegraphics[width=0.9\columnwidth]{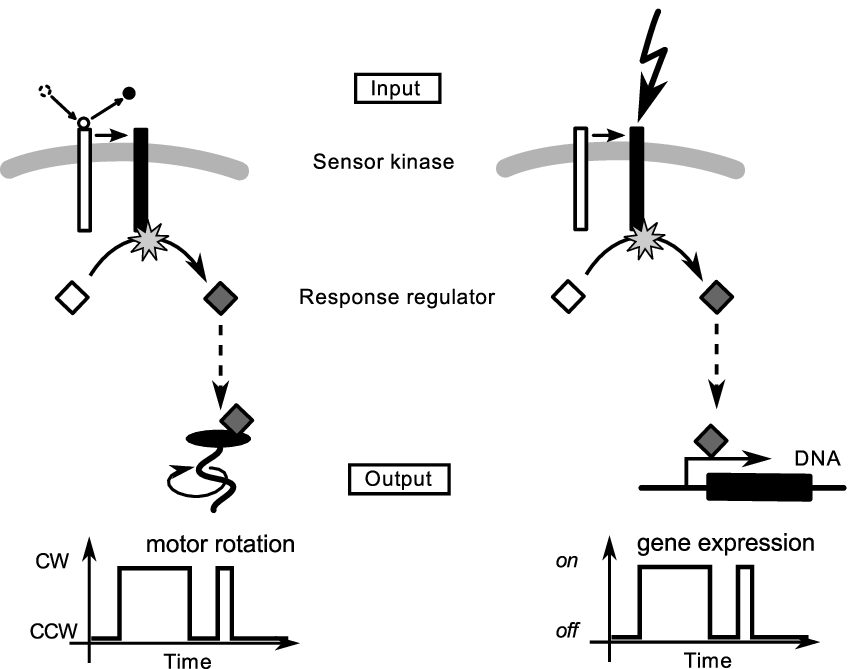}
{
\small
\setlength{\parindent}{1em}

In the chemotaxis pathway ({\it left panel}) and other two-component systems ({\it right panel}), a sensor kinase is activated by a cell-external signal, upon which it autophosphorylates and passes on a phosphoryl group to its response regulator, which typically induces a transcriptional response. The time course of the final output, i.e. gene expression, can be directly mapped onto the binary output of the chemotaxis pathway.

There are particularly well-studied examples of two-component systems: ({\it i}) the VanS (kinase)/VanR (response regulator) system conferring vancomycin resistance in Gram positive bacteria~\citep{HutchButt06}, 
({\it ii}) quorum sensing in {\it Vibrio harveyi}, where the three kinases LuxN, LuxQ and CqsS respond to different autoinducers and first phosphorylate the phosphotransferase LuxO (which has no equivalent in the chemotaxis pathway), which then phosphorylates the response regulator LuxU~\citep{HenBass04}, and 
({\it iii}) the phosphorelay controlling sporulation in {\it Bacillus subtilis}. The relay contains at least four kinases KinA-KinB and the phosphotransferase Spo0F, which phosphorylates the response regulator Spo0A~\citep{JiangHoc00}.

Most other bacterial chemotaxis pathways are more complex than {\it E. coli}'s pathway. For instance, the photosynthetic bacterium {\it Rhodobacter sphaeroides} has several homologues of each of the chemotaxis proteins in {\it E. coli}~\citep{PortWadArm2008}. Interestingly, {\it Rhodobacter} has two chemotaxis receptor clusters, one polar cluster similar to {\it E. coli} and one cytoplasmic cluster, which is thought to sense the metabolic state of the cell. Both clusters need to be present for chemotaxis.
The soil bacterium {\it Bacillus subtilis} has three adaptation systems~\citep{GlekCatOrdal2010}: one based on methylation and demethylation of receptors similar to {\it E. coli} and two independent of receptor methylation, the CheC/CheD system and the CheV system.
Furthermore, in {\it B. subtilis}, sensory adaptation is not determined by the level of receptor methylation but the location of methylation at receptors.
}
\end{multicols}
\end{aBox*}

We predict that the noise power spectrum of the motor reveals a significant low-frequency component when transmitted pathway noise becomes important compared to motor switching noise (Fig.~\ref{fig:motorspec_prediction}) To test this prediction, adaptation rates can be varied using cells expressing different amounts of CheR and CheB from an inducible plasmid. Alternatively, the natural variability in protein expression between cells can be exploited.
Numerous wild-type cells could be monitored. By extracting the adaptation times for chemotactic stimuli, the adaptation rate constants for individual cells can be inferred. Subsequently, the same cells can be used to measure long time traces of motor switching and noise spectra can be calculated.

Our full-pathway model allows us to connect to a large variety of data and literature.
For instance, we also investigated how the response to concentration signals is related to fluctuations in the chemotaxis pathway (Fig.~\ref{fig:FRT}), similar to ideas presented by \citet{ParkPonClu10}. We do not find one unifying fluctuation-response theorem, but a nonlinear trend for the relationship between adaptation times and variance in CheY-P. For small adaptation times we find an approximate linear relationship in line with \citet{ParkPonClu10}.

To describe the noise, the Langevin approximation is expected to work for the phosphorylation and dephosphorylation of the abundant protein CheY. However, its applicability is less clear for receptor signalling due to both extrinsic ligand noise and intrinsic noise from receptor methylation. Furthermore, the switching of the binary motor may constitute relatively large noise. As shown in the {\it SI}, the Master equation and Langevin approximation yield the same results for receptor signalling~\citep{GerCLaEnd10}.
As for the motor, we explicitly tested that the statistical properties of the time series obtained for two-state switching and Langevin equation  are the same. For constant rates, as well as for noisy rates due to fluctuations in CheY-P concentration, the power spectra obtained for the two processes are the same ({\it SI}). In the {\it SI}, we further show that noise terms are indeed sufficiently small that linearisation of the pathway equations is justified.

The bacterial chemotaxis pathway is a member of the large class of two-component systems, containing hundreds of closely related pathways involved in stress response, virulence and inter-cell communication~\citep{LaubGou07,SouArm09}. In these pathways, activation of a sensor histidine kinase results in its autophosphorylation, and subsequently in phosphorylation of a response regulator, which typically binds to DNA and regulates gene expression~(Box~3). The final output, i.e. activation of gene expression, is again binary and hence similar to the bacterial chemotaxis pathway. The analysis presented here may also help elucidate the design of many other pathways and clarify the computational problems cells try to solve.

{\small
\section*{Materials and Methods}

\subsection*{MWC model for activity of receptor complexes}

The MWC model describes signalling by receptor complexes, which can switch between their {\it on} and an {\it off} states. The average activity of a complex is given by
\begin{equation}
 A = \frac{N}{1+e^{F(c, M)}},\label{eq:mwc}
\end{equation}
ranging from zero to $N$. The free-energy difference $F(c, M)$ between the {\it on} and {\it off} state is
\begin{eqnarray}
F(c, M) &=&  N - \frac{1}{2} M + N \left[ \nu_a \ln\left( \frac{ 1+c/K_a^{\text{off}} }{ 1+c/K_a^{\text{off}} } \right) \right.\nonumber\\
&& \left.+ \nu_s \ln\left( \frac{ 1+c/K_s^{\text{off}} }{ 1+c/K_s^{\text{off}} } \right) \right],\label{eq:DF}
\end{eqnarray}
which is a function of the concentration $c$ present at the receptor complex site and the methylation level $M$ of the receptor complex. The methylation level of a complex is the sum of methylation levels of all receptors in a complex.
Here, we consider two receptor types, Tar (indicated by index $a$) with fraction $\nu_a$ of receptors in the complex, and Tsr (indicated by index $s$) with fraction $\nu_s$ of receptors. Receptors are sensitive to attractant MeAsp with dissociation constants $K^\text{on}$ and $K^\text{off}$ in the {\it on} and {\it off} state, respectively.
We use the following parameters for the MWC model for receptor complexes: $K_a^\text{off}=0.02\,\text{mM}$, $K_a^\text{on}=0.5\,\text{mM}$, $K_s^\text{off}=100\,\text{mM}$ and $K_s^\text{on}=10^6\,\text{mM}$~\citep{KeyEndSko06,ClauOleEnd10}.

\subsection*{Noise sources}

\paragraph{Switching noise.}
The switching noise $\eta_{A}(t)$ in Eq.~\eqref{eq:activity} is due to the switching of each receptor complex between {\it on} and {\it off} states. We assume the switching to be a fast process, which can be described by the following dynamics for the the probability of a receptor complex to be {\it on},~$a$:
\begin{equation}
 \frac{da}{dt} = k_1 (N-a) - k_2 a + \eta_{a} (t).\label{eq:microSwitching}
\end{equation}
The noise term $\eta_{a}(t)$ is a Gaussian white noise with zero mean and noise intensity $Q_{a} = 2 k_2 A^*$, where we used that the receptor complex activity $A = \langle a \rangle$ which is equal to the~(quasi)~steady-state activity of $a$, and $A = A^*$ when adapted. 
The power spectrum of $a$ due to switching between {\it on} and {\it off} states is 
\begin{equation}
 S_{a} (\omega) = \frac{Q_{a}}{\omega^2 + (k_1 + k_2)^2},\label{eq:switchspec}
\end{equation}
where $k_1+k_2$ is the characteristic frequency of switching. Hence, the high-frequency component of fluctuations $\delta a(t)$ is reduced due to averaging by the finite rates of switching.
Hence, the power spectrum of activity fluctuations $\eta_{A}(t)$ is
\begin{equation}
 S_{\eta_{A}} (\omega) = \omega^2 S_{a} (\omega). \label{eq:Switchnoise}
\end{equation}

\paragraph{Ligand noise.}

The number of ligand molecules in the vicinity of a receptor complex fluctuates due to binding/unbinding, and potential rebinding of previously bound ligand molecules at this complex, as well as diffusion~\citep{BiaSet05,EndWin09}. Here, we use a simplified description of diffusion to calculate the spectrum of noise in the ligand dynamics $\eta_{c} (t)$ in Eq.~\eqref{eq:ligand}.
Consider a volume whose dimensions are given by the diameter of a receptor complex $s = \sqrt{N} s_R$, where $s_R = 1 \text{ nm}$ is the size of a receptor dimer~\citep{Haz92}. The change of ligand-molecule number $L$ in this volume is determined by the exchange rate $k_D \approx D/ (2s^2)$ due to diffusion~\citep{BergInRandomWalks93}:
\begin{equation}
\frac{dL}{dt} = k_D ( c_0 s^3 - L) + \eta_{L}(t)
\end{equation}
where $k_D L$ is the rate of molecules moving out of the volume by diffusion, and $k_D$ times the mean concentration $c_0$ in solution serves as a proxy of the rate of ligand molecules moving into the volume. The noise term $\eta_{L}(t)$ is assumed to be Gaussian and white, with zero mean and noise intensity $ Q_{L}= D s c_0$.
The power spectrum of the number $L$ and concentration $c$ of molecules at receptor complex $j$ is
\begin{equation}
 S_{L} (\omega) = \frac{D s c_0}{\omega^2 + k_D^2}; \,\,\,  S_{c} (\omega) = \frac{S_{L} (\omega) }{s^6}, \label{eq:Cspec}
\end{equation}
where $s^6$ is the squared volume given by the dimension of the receptor complex.
The zero-frequency limit of the power spectrum of the ligand concentration $S_{c} (0) = c_0 / (D s)$, which corresponds to calculations by Berg and Purcell~\citep{BergPur77} and Bialek and Setayeshgar~\citep{BiaSet05} for the uncertainty in sensing ligand concentration.
The noise $\eta_{c}(t)$ in Eq.~\eqref{eq:ligand} is related to rate of change of the ligand concentration, similar to the considerations of the switching noise above. Hence, the power spectrum of the ligand fluctuations $\eta_{c}(t)$ is 
\begin{equation}
 S_{\eta_{c}} (\omega) = \omega^2 S_{c}(\omega).\label{eq:etaCspec}
\end{equation}

\paragraph{Methylation noise.}
The size of fluctuations in the rate of methylation of a receptor complex $j$ in Eq.~\eqref{eq:methylation}  is estimated from the average rates of methylation and demethylation at the adapted state, respectively. The noise  $\eta_{M}(t)$ is assumed to be Gaussian and white, with zero mean, noise intensity $Q_M = 2 \gamma_R (N - A^*)$ and power spectrum
\begin{equation}
S_{\eta_M}(\omega) = Q_M.  \label{eq:etaMspec}
\end{equation}

\paragraph{Motor switching noise.}
The noise in motor switching rate in Eq.~\eqref{eq:motor} is assumed to be a Gaussian white noise term with zero mean, noise intensity $ Q_{P_\text{CW}}= 2 k_+(A_c^*) (1-P^*_\text{CW})$ and power spectrum
\begin{equation}
S_{\eta_{P_\text{CW}}}(\omega) = Q_{P_\text{CW}}.  \label{eq:motorSwitch}
\end{equation}

\subsection*{Calculation of response functions}
After linearising around the steady state and inserting the Fourier transforms we obtain for the simplified model
\begin{eqnarray}
 -i \omega \Delta \hat{A} & = & - i \omega \frac{\partial A}{\partial M}  \Delta \hat{M} - i \omega \frac{\partial A}{\partial c}  \Delta \hat{c}\label{eq:Ahat}\\
 -i \omega \Delta \hat{M} & = & -\omega_1  \Delta \hat{A}\label{eq:Mhat}\\
 - i \omega \Delta \hat{P}_\text{CW} & = & \omega_2 \Delta \hat{A}_c - \omega_{P_\text{CW}} \Delta \hat{P}_\text{CW}\label{eq:Phat},
\end{eqnarray}
where
\begin{equation}
 \omega_1 = \gamma_R + 3 \gamma_B {{A}^*}^2 = \gamma_R(3-2A_r^*)/A_r^*\label{eq:omega1}
\end{equation}
with ${A}^*=N\cdot A_r^* \approx N/3$~\citep{SouBerg02a} the adapted activity of a receptor complex, $A_r^*$ denoting the adapted activity of individual receptors. In the second equality we have used that at the adapted state $\gamma_R(N-A^*)=\gamma_B{A^*}^3$.
The parameter $\omega_2 = (1-P^*_\text{CW}) \frac{\partial k_+}{\partial A_c} - P^*_\text{CW} \frac{\partial k_-}{\partial A_c}$ is the derivative of the motor switching rates with respect to activity, and $\omega_{P_\text{CW}} = {k_+}^* + {k_-}^*$ is a characteristic frequency due to motor switching at steady state.
$\Delta A$ is the response of every receptor signalling complex, and $\Delta A_c = N_C \Delta A$ is the activity response of all receptor complexes in a cell.
Solving for $\Delta \hat{A}_c$ and $\Delta \hat{P}_\text{CW}$, and division by the stimulus $\Delta \hat{c}$ yields the response functions in Eq.~\eqref{eq:chi_A} and \eqref{eq:chi_motor}.

\subsection*{Calculation of noise power spectra}

To calculate spectra, we linearise the deterministic parts of Eq.~\eqref{eq:activity}-\eqref{eq:motor} similar to the calculation of the response functions, and formally Fourier transform the equations. We obtain
\begin{eqnarray}
 -i \omega \delta \hat{A}_c & = & - i \omega \frac{\partial A}{\partial M} \sum_{j} \delta \hat{M}_j + \frac{\partial A}{\partial c}  \sum_{j} \hat{\eta}_{c_j} \nonumber \\
&&+ \sum_{j} \hat{\eta}_{A_j} \label{eq:Ahat:noise}\\
-i \omega \delta \hat{M}_j & = & -\omega_1  \delta \hat{A}_j + \hat{\eta}_{M_j}\label{eq:Mhat:noise}\\
 - i \omega \delta \hat{P}_\text{CW} & = & \omega_2 \delta \hat{A}_c - \omega_{P_\text{CW}} \delta \hat{P}_\text{CW} + \hat{\eta}_{P_\text{CW}}\label{eq:Phat:noise}.
\end{eqnarray}
We solve for the Fourier transformed activity fluctuations~$\delta \hat{A}_c$ and obtain
\begin{equation}
  \delta \hat{A}_c = \frac{ \frac{\partial A}{\partial M} \sum_{j} \hat{\eta}_{M_j}  + \frac{\partial A}{\partial c}  \sum_{j} {\eta}_{c_j}  + \sum_j \hat{\eta}_{A_j}}{\omega_M - i \omega},
\end{equation}
which yields the power spectrum in Eq.~\eqref{eq:Aspec_simple}. The parameter $\omega_M = \omega_1 \partial A/\partial M$, and we used Eq.~\eqref{eq:Switchnoise} and \eqref{eq:etaCspec}.

From Eq.~\eqref{eq:Phat:noise} we obtain for the Fourier transformed fluctuations in the probability of tumbling mode~$\delta\hat{P}_\text{CW}$
\begin{equation}
  \delta \hat{P}_\text{CW} = \frac{\omega_2\delta \hat{A}_c + \hat{\eta}_{P_\text{CW}}}{\omega_{P_\text{CW}} - i \omega} ,
\end{equation}
and their power spectrum is given by Eq.~\eqref{eq:Pspec_simple}.

\subsection*{Integrated signal response, variance and SNR}

\paragraph{Optimal receptor complex size.}
The integrated response of the receptor activity to a step stimulus is 
\begin{eqnarray}
 \Delta A_c^2&=&\int_{-\infty}^{\infty}d\omega \vert \hat{\chi}_{A_c}(\omega)\Delta \hat{c}(\omega)\vert^2\nonumber\\
&=&\frac{\pi N_C^2\left(\frac{\partial A}{\partial c}\right)^2(\alpha c)^2}{\omega_M},\label{eq:signalA_simple}
\end{eqnarray}
where we inserted Eq.~\eqref{eq:chi_A}. Hence, the activity response scales as $\Delta A_c^2\propto\left(N_\text{tot}/N\right)^2\left (N^2\right)^2/N\propto N$, where we used that $N_C=N_\text{tot}/N$ with $N_\text{tot}$ the total number of receptors in a cell.

The variance of the receptor activity is given by the integral over the power spectrum of activity fluctuations Eq.~\eqref{eq:Aspec_simple}
\begin{eqnarray}
 \langle\delta A_c^2\rangle&=&\frac{N_C}{2\pi}\int_{-\tau^{-1}}^{\tau^{-1}}d\omega \frac{ \omega^2 \left[ S_{a}(\omega) + \left(\frac{\partial A}{\partial c} \right)^2 S_{c}(\omega)\right]}{\omega^2 +\omega_M^2}\nonumber\\
&&+\frac{N_C}{2\pi}\int_{-\tau^{-1}}^{\tau^{-1}}d\omega\frac{\left( \frac{\partial A}{\partial M} \right)^2 Q_M}{\omega^2 +\omega_M^2},
\end{eqnarray}
where we consider the frequency range relevant for motor switching indicated by $\tau^{-1}\approx 0.1\dots 1$ Hz.

The contribution from receptor switching is
\begin{equation}
 \langle\delta A_{c}^2\rangle_a=\frac{N_C}{2\pi}\int_{-\tau^{-1}}^{\tau^{-1}}d\omega \frac{\omega^2 S_a(\omega)}{\omega^2+\omega_M^2}\approx 
\frac{2k_2A_r^*N_\text{tot}}{\pi\tau(k_1+k_2)^2}
\end{equation}
where we used $Q_a$ and inserted Eq.~\eqref{eq:switchspec} for the power spectrum of receptor switching noise and used that it is almost constant and equal to its zero-frequency value over the integration range. Furthermore, the factor $\omega^2/(\omega^2+\omega_M^2)\approx 1$ and $A_r^*=A^*/N$ is the adapted activity of an individual receptor.
Hence, according to this simple calculation the contribution to the variance from receptor switching is roughly constant with receptor complex size.

The contribution from ligand diffusion is
\begin{eqnarray}
 \langle\delta A_{c}^2\rangle_c&=&\frac{N_C}{2\pi}\int_{-\tau^{-1}}^{\tau^{-1}}d\omega \frac{\omega^2\left(\frac{\partial A}{\partial c}\right)^2 S_c(\omega)}{\omega^2+\omega_M^2}\nonumber\\
&\approx& N_C \left(\frac{\partial A}{\partial c}\right)^2\langle\delta c^2\rangle,
\end{eqnarray}
where $\langle\delta c^2\rangle=c_0/(Ds\tau)$ is the variance of the ligand concentration measured during the time interval $\tau$.
We used Eq.~\eqref{eq:Cspec} and the same argument as for the switching noise to calculate the integral.
Hence, the contribution to the variance from the ligand diffusion grows as $\langle\delta A_{c}^2\rangle_c\propto N^3$ as a result of incoherent addition of noise from different receptor complexes and the sensitivity $\partial A/\partial c$ increasing as $N^2$.

The contribution to the variance from receptor methylation is
\begin{eqnarray}
 \langle\delta A_{c}^2\rangle_M&=&\frac{N_C}{2\pi}\left(\frac{\partial A}{\partial M}\right)^2\int_{-\tau^{-1}}^{\tau^{-1}}d\omega \frac{Q_M}{\omega^2+\left(\omega_1 \frac{\partial A}{\partial M}\right)^2}\nonumber\\
&\approx&\frac{2 N_\text{tot}\gamma_R(1-A_r^*)A_r^*}{\omega_1} \frac{\partial A}{\partial M}
\end{eqnarray}
where we defined $\omega_1=\gamma_R+3\gamma_BN^2(A_r^*)^2$, inserted $Q_M=2\gamma_RN(1-A_r^*)$ and $\omega_M=\omega_1(\partial A/\partial M)$.
Hence, $\langle\delta A_{c}^2\rangle_M$ grows approximately linearly with receptor complex size.

The SNR grows linearly with $N$ for small complex sizes, and decreases as $N^{-2}$ for larger complex sizes, resulting in an optimal medium receptor complex size, in qualitative agreement with Fig.~\ref{fig:SNR_vary_N}.

\paragraph{Optimal adaptation rates.}
The integrated signal response of the receptor activity Eq.~\eqref{eq:signalA_simple}
\begin{equation}
 \Delta A_c^2=\frac{\pi N_C^2\left(\frac{\partial A}{\partial c}\right)^2(\alpha c)^2}{\omega_1\frac{\partial A}{\partial M}},
\end{equation}
where the numerator expresses the initial response of receptors of concentration changes and the denominator the filtering by adaptation.
The sensitivity $\partial A/\partial c=NA_r^*(1-A_r^*)h(c)$, where $h(c)=\partial F/\partial c$, $\omega_1=\gamma_R+3\gamma_BN^2(A_r^*)^2$, and $\partial A/\partial M=NA_r^*(1-A_r^*)/2$. The adapted activity can be obtained analytically for our simplified model from the steady state of the methylation dynamics Eq.~\eqref{eq:methylation},
\begin{equation}
 {A_r}^* = \sqrt[3]{\frac{1}{2}\beta + \sqrt{\frac{\beta^2}{4} + \frac{\beta^3}{27}}} - \frac{\beta}{3 \sqrt[3]{\frac{1}{2}\beta + \sqrt{\frac{\beta^2}{4} + \frac{\beta^3}{27}}}},
\end{equation}
and is only a function of the ratio $\beta=\gamma_R/\gamma_B$. Expanding the adapted activity around $A_r^*=0$ (for $\gamma_R\to 0$) yields $A_r^*\propto \gamma_R^{1/3}$, and around $A_r^*=1$ (for $\gamma_B\to 0$) yields $A_r^*\propto \gamma_B$. Similarly, $\omega_1\propto\gamma_R^{2/3}$ (const.+$\gamma_B^{4/3}$). Hence, $\partial A/\partial c\propto\gamma_R^{1/3}$ ($\gamma_B$) and $\omega_1 \partial A/\partial M\propto\gamma_R^{4/3}$ ($\gamma_B$).

The initial response to concentration changes decreases slower than adaptation times, resulting in an increased signal response for vanishing $\gamma_R$. For vanishing $\gamma_B$, the initial response to concentration changes decreases faster than adaptation speed, hence yielding a vanishing signal response.
The overall dependence of the integrated signal response is $\Delta A_c^2\propto \gamma_R^{-1/3}$ ($\gamma_B$) for $\gamma_R\to 0 (\gamma_B\to 0)$.
For the contributions to the variance of the receptor activity from receptor switching, ligand diffusion and receptor methylation dynamics we obtain $\langle\delta A_c^2\rangle_a \propto\gamma_R^{1/3}$ ($\gamma_B$), $\langle\delta A_c^2\rangle_c\propto \gamma_R^{2/3}$ ($\gamma_B^2$) and $\langle\delta A_c^2\rangle_M\propto\gamma_R$ ($\gamma_B^{2/3}$), respectively.

Hence, according to our simplified model the SNR of the receptor activity goes as SNR $\propto\gamma_R^{-2/3}$ ($\gamma_B^{4/3}$), in qualitative agreement with Fig.~\ref{fig:SNR_vary_RB}.
} 

\section*{Acknowledgements}
We thank Richard Berry, Martin Buck, Tom Duke and William Ryu for helpful discussions. RGE was supported by Biotechnological and Biological Sciences Research Council grant BB/G000131/1 and the Centre for Integrative Systems Biology at Imperial College (CISBIC).

\section*{Author contributions}

DC and RGE conceived and designed the study, performed analytical calculations, analysed the data, and wrote the
paper. DC performed computer simulations.


\end{document}


\title{Supplementary information:\\Noise characteristics of the {\it Escherichia coli} rotary motor}
\author{Diana Clausznitzer$^{1,2,3}$ and Robert G. Endres$^{1,2, *}$}
\date{$^1$ Division of Molecular Biosciences, Imperial College London, SW7 2AZ London, UK;\\
$^2$ Centre for Integrative Systems Biology at Imperial College, Imperial College London, SW7 2AZ London, UK;\\
$^3$ BioQuant, Universit\"at Heidelberg, 69120 Heidelberg, Germany\\
$^*$e-mail: r.endres@imperial.ac.uk}

\maketitle
\tableofcontents


\section{Stochastic differential equations}

In the main text, we presented a simplified model of the chemotaxis pathway to illustrate signalling and noise transmission. Here, we discuss a model for the full signalling pathway and present the response functions and noise spectra.
Equation~(1) in the main text describes the total signalling activity $A_c$ of all receptor complexes in a cell in response to changes in the methylation level of the complexes and ligand concentration is given by
\begin{equation}
\frac{dA_c}{dt} = \sum_{j=1}^{N_c} \frac{\partial A}{\partial M} \frac{dM_j}{dt} + \frac{\partial A}{\partial c} \frac{dc_j}{dt} + \eta_{A_j} (t).\label{eq:fullpathway:activity}
\end{equation}
The dynamics of the methylation level of complex $j$ is described by 
\begin{eqnarray}
 \frac{dM_j}{dt} & = &  \gamma_R (N-A_j) - \gamma_B A_j B_p^2 + \eta_{M_j} (t)\\
	& = & \gamma_R (N-A_j) - \frac{\gamma_B }{V_\text{cell}^2}A_j N_{B_p}^2 + \eta_{M_j} (t)
\end{eqnarray}
Note that here we explicitly include the number of CheB-P  (${B_p}$) molecules $N_{B_p}$ in the demethylation term, with $V_\text{cell}$ the cell volume~(cf.~Eq.~(3) in the main text). We denote the activity of complex $j$ by $A_j$.
The dynamics of the concentration according to Eq.~(2) in the main text is
\begin{equation}
\frac{dc_j}{dt} = \frac{d\langle c\rangle}{dt} + \eta_{c_j}(t).
\end{equation}
In addition, we take into account phosphorylation and dephosphorylation of CheA (${A_p}$), CheY (${Y_p}$) and CheB (${B_p}$), which are described by the following equations:
\begin{eqnarray}
 \frac{dN_{A_p}}{dt} & = & A_{c} \left( \frac{k_A}{N_c N}\right) ( N_{A, \text{tot}} - N_{A_p} ) - \left( \frac{k_y}{V_{\text{cell}}} \right) ( N_{Y, \text{tot}} - N_{Y_p} ) N_{A_p} +\nonumber\\
&& - \left( \frac{k_b}{V_{\text{cell}}} \right) ( N_{B, \text{tot}} - N_{B_p} ) N_{A_p} + \eta_{A, p} (t) + \eta_{A, Y_p} (t) + \eta_{A, B_p} (t)\\
\frac{dN_{Y_p}}{dt} & = & \left( \frac{k_y}{V_{\text{cell}}} \right) ( N_{Y, \text{tot}} - N_{Y_p} ) N_{A_p} - k_{-y} N_{Y_p} - \eta_{A, Y_p} (t) + \eta_{-Y_p} (t)\\
\frac{dN_{B_p}}{dt} & = & \left( \frac{k_b}{V_{\text{cell}}} \right) ( N_{B, \text{tot}} - N_{B_p} ) N_{A_p} - k_{-b} N_{B_p} - \eta_{A, B_p} (t) + \eta_{-B_p} (t)
\end{eqnarray}
with $N_i$ the number of molecules of species $i$ in a cell volume. Note, that we neglected the binding of CheY-P to its phosphatase CheZ  for simplicity, and describe dephosphorylation of CheY-P by the effective dephosphorylation rate $k_{-y} N_{Y_p}$. As shown in Sec.~\ref{app:highfrequfilter}, this simplification has no qualitative effect on the response function.
The term $\eta_{A, p}$ describes the noise associated with CheA autophosphorylation.  The terms $\eta_{A, B_p}(t)$ and $\eta_{A, Y_p}(t)$ represent the noise generated in phosphorylation of CheB and CheY by CheA, respectively. The terms $\eta_{-B_p}(t)$ and $\eta_{-Y_p}(t)$ describe the noise associated with dephosphorylation. Note that some noise terms appear in two equations. This is due to the fact that we assign noise terms to a specific process, e.g. phosphorylation of CheY by CheA. Hence, the corresponding noise term $\eta_{A, Y_p} (t)$ appears in the dynamics of CheA-P and of CheY-P. As a positive fluctuation in the dynamics of CheA-P due to phosphorylation of CheY corresponds to a negative fluctuation in the dynamics of CheY-P, these noise terms appear with opposite signs in the two equations.
The noise intensities and parameter values of the model are summarised in Sec.~\ref{sec:app:parametersFullModel}.
Finally, the dynamics of the motor is described in terms of the probability of the tumbling mode $P_\text{CW}$
\begin{equation}
\frac{dP_\text{CW}}{dt}		= k_+(N_{Y_p}) (1-P_\text{CW}) - k_-(N_{Y_p}) P_\text{CW} + \eta_{P_\text{CW}} (t),\label{eq:fullpathway:motor}
\end{equation}
where we use experimentally derived switching rates $k_+$ and $k_-$ as a function of CheY-P concentration (cf. next section and Fig.~2C in the main text).

\section{Model for motor switching\label{app:switchrates}}

\citet{TurnSamBerg99} presented a model for motor switching to explain the observed motor switching rates.
The model for motor switching is an MWC model, where 26 subunits of the motor assume one of two states corresponding to CW and CCW rotation. While these subunits bind the molecule CheY-P independently of each other, the switching of states occurs cooperatively. The authors derive the overall rates of switching (averaging over all possible CheY-P occupancy states) as
\begin{eqnarray}
 k_+ (Y_p)  &= & k_+(0) \cdot \exp\bigg\{ m_\text{coop} \cdot \ln\left[ \left(1+\frac{\mu Y_p}{K_{\text{CCW}}}\right)\bigg/\left( 1+\frac{Y_p}{K_{\text{CCW}}}\right)\right]\bigg\}\\
 k_- (Y_p)  &= & k_-(0) \cdot \exp\bigg\{ {m_\text{coop} \cdot \ln\left[ \left({1+\frac{\mu Y_p}{K_{\text{CCW}}}}\right)\bigg/\left({1+\frac{Y_p}{K_{\text{CW}}}}\right)\right]}\bigg\},
\end{eqnarray}
where $m_\text{coop}=26$ is the number of motor subunits, $-k_B T\ln(\mu)$ is the free-energy difference of switching per molecule of CheY-P, and $K_{\text{CCW}}$ and $K_{\text{CW}}$ are the dissociation constants for binding CheY-P in the CCW and CW state, respectively.

Motor switching rates $k_+$ and $k_-$ have been derived experimentally as a function of the concentration of a signalling mutant CheY$^{**}$, which is constitutively active~\citep{TurnSamBerg99}, as shown in Fig.~2C in the main text. To obtain the switching rates in terms of CheY-P, rather than signalling mutant CheY$^{**}$, we rescaled the dissociation constants of CheY binding to the motor such that the switching rates are equal, i.e. CW bias about 1/2, at CheY-P concentration 3.2 $\mu M$~\citep{CluSurLei00}. We fitted the above model to the experimental data in Fig.~2C and the CW bias at 33$^{\circ}$~\citep{TurnSamBerg99} and used the rate constants $k_+$ and $k_-$ in our full pathway model (Eq.~\eqref{eq:fullpathway:motor}).

\section{Linearisation of the model}

Similar to the presentation for the simplified model in the main text, we linearise Eq.~\eqref{eq:fullpathway:activity}-\eqref{eq:fullpathway:motor}
and insert the Fourier transforms of the dynamical variables to obtain the Fourier transformed response functions $\hat{\chi}_R(\omega)$ and noise spectra $S_R(\omega)$ for the signalling pathway.
The linearised equations read 
\begin{eqnarray}
\!\!\!\frac{d(\delta A_c)}{dt}  \!\!\!& = \!\!\!& - \sum_j \frac{\partial A}{\partial M} \left( \lambda_1 \delta A_j + \lambda_9 \delta N_{B_p} \right) + \frac{\partial A}{\partial c} \frac{d (\delta c_j)}{dt} + \frac{\partial A}{\partial M} \eta_{M_j} (t) + \eta_{A_j} (t)\label{eq:fullpathway:activity:linearized}\\
\!\!\!\frac{d(\delta N_{A_p})}{dt}  \!\!\!& = \!\!\!& \lambda_2 \delta A_c - \lambda_3 \delta N_{A_p} + \lambda_4 \delta N_{Y_p} + \lambda_{10} \delta N_{B_p} + \eta_{A_p} (t) + \eta_{A, Y_p} (t) + \eta_{A, B_p} (t)\\
\!\!\!\frac{d(\delta N_{Y_p})}{dt}  \!\!\!& = \!\!\!& \lambda_5 \delta N_{A_p} - \lambda_6 \delta N_{Y_p} - \eta_{A, Y_p} (t) + \eta_{-Y_p} (t)\\ 
\!\!\!\frac{d(\delta N_{B_p}) }{dt} \!\!\!& = \!\!\!& \lambda_{11} \delta N_{A_p} - \lambda_{12} \delta N_{B_p} - \eta_{A, B_p} (t) + \eta_{-B_p} (t)\\ 
\!\!\!\frac{d(\delta P_\text{CW})}{dt}  \!\!\!& = \!\!\!& \lambda_7 \delta N_{Y_p} - \lambda_8 \delta P_\text{CW} + \eta_{P_\text{CW}} (t)\label{eq:fullpathway:motor:linearized}
\end{eqnarray}
with rate constants of the linearised model given in Table~\ref{tab:linearizedrateconstants}.

\section{Response functions}

The response functions can be calculated from the linearised Eq.~\eqref{eq:fullpathway:activity:linearized}-\eqref{eq:fullpathway:motor:linearized} without noise after inserting the Fourier transforms of the dynamical variables. The Fourier transformed response functions of CheA-P, CheY-P and the motor are
\begin{eqnarray}
\hat{\chi}_{A_c} (\omega) & = & \frac{ -i\omega  N_c \frac{\partial A}{\partial c}- \lambda_9  N_c \frac{\partial A}{\partial M} \hat{\chi}_{N_{B_p}}(\omega) }{ \lambda_1 \frac{\partial A}{\partial M}  -i\omega} \\
\hat{\chi}_{N_{A_p}}(\omega) & = & 
\left( -i\omega \lambda_2 N_c \frac{\partial A}{dc} \left(\lambda_6-i\omega\right)\left(\lambda_{12}-i\omega\right) \right) \cdot \nonumber\\
&& \left\lbrace \left(\lambda_1 \frac{\partial A}{\partial M}-i\omega\right) \cdot \left[ \left(\lambda_3-i\omega\right)\left(\lambda_6-i\omega\right)\left(\lambda_{12}-i\omega\right) - \lambda_{10} \lambda_{11} \left(\lambda_6-i\omega\right) + \right.\right.\nonumber \\
&& \left.\left. - \lambda_4 \lambda_5 \left(\lambda_{12}-i\omega\right)  \right] + \lambda_2 \lambda_9 \lambda_{11} N_c \frac{\partial A}{\partial M}\left(\lambda_6-i\omega\right) \right\rbrace^{-1} \\
\hat{\chi}_{N_{Y_p}}(\omega) & = & \frac{\lambda_5  }{ \lambda_6 -i\omega } \hat{\chi}_{N_a}(\omega)\label{eq:Yresponse}\\
\hat{\chi}_{P_\text{CW}}(\omega) & = & \frac{ \lambda_7 }{ \lambda_8-i\omega } \hat{\chi}_{N_y}(\omega)\\
\hat{\chi}_{N_{B_p}}(\omega) & = & \frac{\lambda_{11} }{  \lambda_{12}-i\omega  }  \hat{\chi}_{N_{A_p}}.
\end{eqnarray}
From these equations we observe that CheA-P, CheY-P and the motor are in a cascade. At each level of the cascade, a new filter proportional to $(\lambda_i -i\omega)^{-1}$ is introduced which simply multiplies the response function of the previous level of the cascade. The characteristic frequencies $\lambda_i$ contain the forward and backward rates of the relevant processes.

\section{Noise spectra}
The noise spectra can be calculated from the linearised Eq.~\eqref{eq:fullpathway:activity:linearized}-\eqref{eq:fullpathway:motor:linearized}. After inserting the Fourier transforms of the dynamical variables, calculating the absolute squared value and averaging, we obtain the noise spectra for CheA-P, CheY-P and the motor shown in Fig.~3 in the main text

\begin{eqnarray}
S_{A_c} (\omega) & = &  \left( \left\vert (\lambda_3-i\omega)(\lambda_6-i\omega)(\lambda_{12}-i\omega) - \lambda_4\lambda_5(\lambda_{12}-i\omega) - \lambda_{10}\lambda_{11}(\lambda_6-i\omega) \right\vert^2   \cdot \right.\nonumber\\
&&  \hspace{2em} N_c \left[\omega^2 \left( \frac{\partial A}{\partial c} \right)^2 S_c(\omega) + \omega^2 \left( S_a(\omega) +Q_M  \right)\right]+\nonumber\\
&& + \left\vert -\lambda_9 N_c \frac{\partial A}{\partial M} \left[ (\lambda_3-i\omega)(\lambda_6-i\omega)-\lambda_4\lambda_5 \right] \right\vert^2  Q_{-B_p} +\nonumber\\
&& + \left\vert \lambda_9\lambda_{11} N_c \frac{\partial A}{\partial M}(\lambda_6-i\omega) \right\vert^2 Q_{A_p} + \left\vert \lambda_4\lambda_9\lambda_{11} N_c \frac{\partial A}{\partial M} \right\vert^2 Q_{-Y_p} + \nonumber\\
&& + \left\vert \lambda_9 N_c \frac{\partial A}{\partial M} \left[  \lambda_4\lambda_5+\lambda_11(\lambda_6-i\omega) - (\lambda_3-i\omega)(\lambda_6-i\omega) \right] \right\vert^2 Q_{A,B_p} + \nonumber\\ 
&& \left. +  \left\vert \lambda_9\lambda_{11} N_c \frac{\partial A}{\partial M}\left[(\lambda_6-i\omega)-\lambda_4\right] \right\vert^2 Q_{A,Y_p}\right) \cdot\nonumber\\
&& \hspace{2em} \left\vert \left(\lambda_1\frac{\partial A}{\partial M}-i\omega\right)\left[  (\lambda_3-i\omega)(\lambda_6-i\omega)(\lambda_{12}-i\omega) -\lambda_{10}\lambda_{11}(\lambda_6-i\omega) + \right.\right.\nonumber\\
&& \left.\left. \hspace{8em} - \lambda_4\lambda_5(\lambda_{12}-i\omega)  \right] + \lambda_2\lambda_9\lambda_{11} N_c \frac{\partial A}{\partial M}(\lambda_6-i\omega)\right\vert^{-2}
\end{eqnarray}

\begin{eqnarray} 
S_{N_{A_p}}(\omega) & = & 
\left( N_c \lambda_2^2 \omega^2 \left( \frac{\partial A}{dc} \right)^2 \left\vert \left(\lambda_6-i\omega\right)\left(\lambda_{12}-i\omega\right) \right\vert^2 S_c(\omega) + \right.\nonumber\\
&& + \left\vert \lambda_2 (\lambda_6-i\omega)(\lambda_{12}-i\omega) \right\vert^2 N_c \left( \omega^2 S_a(\omega) + Q_M \right) + \nonumber\\
&& + \left\vert \left(\lambda_1\frac{\partial A}{\partial M} -i\omega\right)(\lambda_6-i\omega)(\lambda_{12}-i\omega)  \right\vert^2 Q_{A_p} +\nonumber\\
&&  + \left\vert (\lambda_6-i\omega)\left(-\lambda_2 \lambda_9 N_c \frac{\partial A}{\partial M} + \lambda_{10} \left(\lambda_1\frac{\partial A}{\partial M}-i\omega\right) \right) \right\vert^2 Q_{-B_p} +\nonumber\\
&& + \left\vert \lambda_4 \left(\lambda_1\frac{\partial A}{\partial M} -i\omega\right)(\lambda_{12}-i\omega) \right\vert^2 Q_{-Y_p} + \nonumber\\
&& +\left\vert  (\lambda_6-i\omega)(\lambda_{12}-i\omega)(\lambda_1\frac{\partial A}{\partial M}-i\omega) +(\lambda_6-i\omega)(\lambda_2 \lambda_9 N_c \frac{\partial A}{\partial M}  + \right.\nonumber\\
&& \left. \hspace{10em}- \lambda_{10} \left(\lambda_1\frac{\partial A}{\partial M}-i\omega\right) ) \right\vert^2 Q_{A, B_p} + \nonumber\\
&& \left. + \left\vert \left(\lambda_1\frac{\partial A}{\partial M}-i\omega\right)(\lambda_6-i\omega)(\lambda_{12}-i\omega) - \lambda_4 \left(\lambda_1\frac{\partial A}{\partial M}-i\omega\right)(\lambda_{12}-i\omega) \right\vert^2 Q_{A, Y_p} \right) \cdot \nonumber\\
&& \hspace{2em}\left\vert \left(\lambda_1 \frac{\partial A}{\partial M}-i\omega\right) \cdot \left[ \left(\lambda_3-i\omega\right)\left(\lambda_6-i\omega\right)\left(\lambda_{12}-i\omega\right) - \lambda_{10} \lambda_{11} \left(\lambda_6-i\omega\right) + \right.\right.\nonumber \\
&& \hspace{2em}\left.\left. - \lambda_4 \lambda_5 \left(\lambda_{12}-i\omega\right)  \right] + \lambda_2 \lambda_9 \lambda_{11} N_c \frac{\partial A}{\partial M} \left(\lambda_6-i\omega\right) \right\vert^{-2}
\end{eqnarray}
\begin{eqnarray}
S_{N_{Y_p}}(\omega) & = & \frac{\lambda_5^2(\omega) }{ \left\vert\lambda_6 -i\omega \right\vert^2} \cdot \left( N_c \lambda_2^2 \omega^2 \left(\frac{\partial A}{dc}\right)^2 \left\vert \left(\lambda_6-i\omega\right)\left(\lambda_{12}-i\omega\right) \right\vert^2 S_c(\omega) + \right.\nonumber\\
&& + \left\vert \lambda_2 (\lambda_6-i\omega)(\lambda_{12}-i\omega) \right\vert^2 N_c \left( \omega^2 S_a(\omega) + Q_M \right) + \nonumber\\
&& + \left\vert \left(\lambda_1\frac{\partial A}{\partial M} -i\omega\right)(\lambda_6-i\omega)(\lambda_{12}-i\omega)  \right\vert^2 Q_{A_p} + \nonumber\\
&&  + \left\vert (\lambda_6-i\omega)\left(-\lambda_2 \lambda_9 N_c \frac{\partial A}{\partial M} + \lambda_{10} \left(\lambda_1\frac{\partial A}{\partial M}-i\omega\right) \right) \right\vert^2 Q_{-B_p} + \nonumber\\
&& + \left\vert \frac{\lambda_5}{\lambda_6 -i\omega} \left[ \lambda_4 \left(\lambda_1\frac{\partial A}{\partial M} -i\omega\right)(\lambda_{12}-i\omega) \right] + 1 \right\vert^2 Q_{-Y_p} + \nonumber\\
&& +\left\vert  (\lambda_6-i\omega)(\lambda_{12}-i\omega)(\lambda_1\frac{\partial A}{\partial M}-i\omega) +(\lambda_6-i\omega)(\lambda_2 \lambda_9 N_c \frac{\partial A}{\partial M}  + \right.\nonumber\\
&& \left. \hspace{10em}- \lambda_{10} \left(\lambda_1\frac{\partial A}{\partial M}-i\omega\right) ) \right\vert^2 Q_{A, B_p} +\nonumber\\
&& + \left\vert \frac{\lambda_5}{\lambda_6 -i\omega} \left[ \left(\lambda_1\frac{\partial A}{\partial M}-i\omega\right)(\lambda_6-i\omega)(\lambda_{12}-i\omega) + \right.\right. \nonumber\\
&& \left.\left.\left.\hspace{10em} - \lambda_4 \left(\lambda_1\frac{\partial A}{\partial M}-i\omega\right)(\lambda_{12}-i\omega) \right] -1 \right\vert^2 Q_{A, Y_p} \right)\cdot     \nonumber\\
&& \hspace{2em}\left\vert \left(\lambda_1 \frac{\partial A}{\partial M}-i\omega\right)\cdot \left[ \left(\lambda_3-i\omega\right)\left(\lambda_6-i\omega\right)\left(\lambda_{12}-i\omega\right) - \lambda_{10} \lambda_{11} \left(\lambda_6-i\omega\right) + \right.\right.\nonumber \\
&& \hspace{2em}\left.\left. - \lambda_4 \lambda_5 \left(\lambda_{12}-i\omega\right)  \right] + \lambda_2 \lambda_9 \lambda_{11} N_c \frac{\partial A}{\partial M}\left(\lambda_6-i\omega\right) \right\vert^{-2} \\
S_{P_\text{CW}}(\omega) & = & \frac{ \lambda_7^2 S_{N_{Y_p}}(\omega) + Q_{P_\text{CW}}}{ \left\vert\lambda_8-i\omega \right\vert^2}
%
\end{eqnarray}

\section{Number of high-frequency filters\label{app:highfrequfilter}}
In Fig.~2A in the main text it is apparent that our model does not fully reproduce the high-frequency response. The high-frequency response seems to be a third-order filter in the frequency range shown, while our model only produces a second-order filter due to CheY-P and motor-switching dynamics. A third filter due to the autophosphorylation dynamics, which is included in our model, becomes only relevant at higher frequencies.
Here, we discuss where an additional filter could originate.

We explicitly consider the CheY-P/CheZ binding step, and write down the equations for the dynamics of the concentration of CheY-P, denoted by $y$, and of CheY-P/CheZ complex, $yz$,
\begin{eqnarray}
\frac{d{y}}{dt} 	& = & g_{Y} \left( Y_{\text{tot}}-y \right) a - g_{1} \left(Z_{\text{tot}}-yz\right) y + g_2 yz\\
 \frac{d({yz})}{dt}	& = & g_1 (Z_{\text{tot}}-yz) y - \left(g_2+g_3\right) yz, 
\end{eqnarray}
where $a$ is the concentration of phosphorylated CheA and $g_i$ are the rates of phosphorylation of CheY ($g_Y$), CheY-P/CheZ complex formation ($g_1$), dissociation of CheY-P/CheZ complexes ($g_2$) and CheY-P dephosphorylation ($g_3$).
Linearising around the steady state~$(a^*, y^*, yz^*)$ yields
\begin{eqnarray}
\frac{d (\Delta{y})}{dt} 	& = & \underbrace{g_{Y} \left( Y_{\text{tot}}-y^* \right)}_{\tilde\lambda_1} \Delta a - \underbrace{\left [ g_Y a^* + g_1 \left(Z_{\text{tot}}-yz^*\right) \right]}_{\tilde\lambda_2} \Delta y + \underbrace{\left( g_1 y^*  + g_2\right)}_{\tilde\lambda_3} \Delta yz\\
 \frac{d (\Delta{yz})}{dt}	& = & - \underbrace{\left[ g_1 y^* + g_2 + g_3 \right]}_{\tilde\lambda_3 + g_3 } \Delta yz + \underbrace{g_1 \left(Z_{\text{tot}}-yz^*\right)}_{\tilde\lambda_4} \Delta y.
\end{eqnarray}
Hence, we obtain for the Fourier transform of deviations in the CheY-P concentration 
\begin{equation}
\Delta \hat{y} = \frac{\tilde\lambda_1 (-i\omega + \tilde\lambda_3 + g_3)}{(-i\omega+\tilde\lambda_2) (-i\omega+\tilde\lambda_3+g_3) - \tilde\lambda_3\tilde\lambda_4} \Delta\hat{a}.
\end{equation}
To make the analysis easier, we can factorise the polynomial in the denominator,
\begin{equation}
\Delta \hat{y} = \frac{\tilde\lambda_1 (-i\omega + \tilde\lambda_3 + g_3)}{(-i\omega+a_1) (-i\omega+a_2)} \Delta\hat{a},\label{eq:Yresponse_YZ}
\end{equation}
with $a_{1, 2} =  (\tilde\lambda_3 + g_3+\tilde\lambda_2)/2 \pm \sqrt{(\tilde\lambda_3 + g_3+\tilde\lambda_2)^2/4 - \tilde\lambda_2 (\tilde\lambda_3+g_3) + \tilde\lambda_3\tilde\lambda_4}$.

We are interested in the behaviour of the frequency-dependent prefactor. Specifically, we ask if by considering the CheY-P/CheZ complex formation we obtain an additional high-frequency filter compared to Eq.~\eqref{eq:Yresponse}.
It is obvious from Eq.~\eqref{eq:Yresponse_YZ} that under most parameter combinations we obtain $1/\omega$ behaviour at high frequencies. Hence, no additional filter is introduced.
A special case appears for $\tilde\lambda_3+k3 \gg a_1, \, a_2$. In this case, a $1/\omega^2$ behaviour is observed for medium frequencies $\max(a_1, \, a_2) \gg \omega \gg \tilde\lambda_3+g_3$. Hence, additional filter appears. At high frequencies $\omega > \tilde\lambda_3+g_3$, the prefactor has $1/\omega$ behaviour.
However, analysing the expressions for $a_1$ and $a_2$ reveals that $\max(a_1, \, a_2)$ is always greater or equal to $(\tilde\lambda_3 + g_3+\tilde\lambda_2)/2$. Therefore, this case does not occur for our dynamics.
In conclusion, considering CheY-P/CheZ complex formation does not introduce an additional high-frequency filter.

Other processes in the signalling pathway neglected here are the oligomerisation of CheY-P/CheZ complexes~\citep{EisenbachInEisenbach04, BlatEis1996a,BlatEis1996b} and a potential slow release of CheY-P from the sensory complex as discussed by \citet{BlatGillEis98}.
%
Oligomerisation of CheY-P/CheZ complexes for efficient dephosphorylation is similar to CheY-P/CheZ complex formation considered above, and by a similar discussion does not introduce an additional high-frequency filter.
However, a delayed release of CheY-P represents effectively a step between CheY phosphorylation and motor switching in the signalling pathway, and hence could introduce a relevant filter if the process is sufficiently slow.
Another possibility to explain the steep frequency-dependence of the response function is that the duration of experimental pulses was long enough to leave a signature.

\section{Alternative Master-equation approach\label{app:masterequationapproach}}
Alternative to the Langevin approach, which assumes small fluctuations, we can write down a Master equation for the pathway.
Here, we focus on ligand and methylation dynamics at the receptor cluster. Each state of the pathway is described by the variables [$L_j$, $M_j$] at each of the receptor complexes, where $L_j=c_j s^3$ is the number of molecules in a small volume at the receptor complex and $M_j$ the total methylation level of the receptor complex.
Assuming for simplicity only one receptor complex, the Master equation for the probability density $p$ is
\begin{eqnarray}
 \frac{\partial p(L, M, t)}{\partial t} & = & k_D (s^3 c_0) p(L-1, M, t)\nonumber\\
&& + k_D (L+1) p(L+1, M, t)\nonumber\\
&& + \gamma_R [1-A(L, M-1)] p(L, M-1, t)\nonumber\\
&& + \gamma_B [A(L, M+1)]^3 p(L, M+1, t)\nonumber\\
&& - \{k_D (s^3 c_0+L) + k_D (s^3 c_0) + \gamma_R [1-A(L, M)] + \gamma_B [A(L, M)]^3 \} p(L, M, t)
\end{eqnarray}
For small noise, using van Kampen’s $\Omega$ expansion~\citep{vanKampen2007} for the variances of fluctuations in the number of ligand molecules and the methylation level at steady state~\citep{GerCLaEnd10}, we obtain 
\begin{eqnarray}
\langle\delta L^2\rangle&=&c_0 s^3 \label{eq:master_L}\\
\langle\delta M^2\rangle&=&\frac{1}{(3-2 A^*)\beta}+\frac{\gamma_R(3-2A^*)\left(\frac{\partial A}{\partial c}\right)^2c}{{A^*}^2(1-A^*)[k_D+
\gamma_R(3-2A^*)(1-A^*)\beta]\beta}\label{eq:master_M},
\end{eqnarray}
with $\beta=1/2$ the free-energy difference due to adding one methyl group (in units of $k_BT$). The first term is due to fluctuations in the rate of methylation and demethylation and the second term is due to transmitted fluctuations in the activity from ligand noise. 
This corresponds to the results from the Langevin approach. Specifically, for ligand fluctuation we obtain the same variance after integration of the power spectrum Eq.~(17) in the main text.
Furthermore, the power spectrum of the receptor complex methylation level using the simplified model from the main text is 
\begin{equation}
 S_M(\omega) = \frac{Q_M + (\gamma_R + 3 \gamma_B{A^*}^2)^2 \left( \frac{\partial A}{\partial c}\right)^2 S_c(\omega)}{\omega^2 + \omega_M^2}.
\end{equation}
The variance of the methylation level, obtained by integration of the power spectrum, corresponds to the above result.

\section{Test of Langevin approximation for motor dynamics\label{app:langevin2state}}
We chose to describe the dynamics of the motor using the Langevin Eq.~(4) in the main text
\begin{equation}
\frac{dP_\text{CW}}{dt}		= k_+ (1-P_\text{CW}) - k_- P_\text{CW} + \eta_{P_\text{CW}} (t) 
\end{equation}
with switching rates from CCW to CW (first term) and from CW to CCW (second term), as well as an additive Gaussian white noise term (last term) with zero mean and autocorrelation $\langle \eta_{P_\text{CW}}(t) \eta_{P_\text{CW}}(t') \rangle = Q_{P_\text{CW}} \delta(t-t')$ with $Q_{P_\text{CW}}=2 k_+ (1-P^*_t)=2k_+k_-/(k_++k_-)$. For constant switching rate constants $k_+$ and $k_-$, the power spectrum $P_\text{CW}$ is (cf. Eq.~(9) in the main text)
\begin{equation}
 S_{P_\text{CW}}(\omega)=\frac{Q_{P_\text{CW}}}{\omega^2+(k_++k_-)^2}.
\end{equation}
To see that this is a valid description of the binary motor-switching process, we calculate the spectrum exactly according to the derivation by \citet{chapter6INStr67}. For a stochastic two-state process, whose time interval lengths in each of the two states $\tau_1$ and $\tau_2$, respectively, are independent and identically distributed random variables, the power spectrum is given in terms of the Fourier transforms of the waiting time distributions $\Theta_1(\omega)$ and $\Theta_2(\omega)$ for each of the states,
\begin{equation}
 S(\omega) = \frac{2}{\omega^2 (\langle \tau_1 \rangle + \langle \tau_2 \rangle)} \Re{ \frac{ [1-\Theta_1(\omega) ] [1-\Theta_2(\omega) ] }{ 1- \Theta_1(\omega)\Theta_2(\omega) }  },
\end{equation}
where $\Re$ indicates the real part. Assuming for the motor that switching between the states CW and CCW, respectively, follows exponential interval distributions determined by rates $k_+$ and $k_-$~\citep{BloSegBerg83,BaiBranchBerry2010}, the Fourier transforms of the waiting time distributions are given by,
\begin{eqnarray}
 \Theta_{CW}(\omega) & = & \frac{k_+}{k_+ - i \omega}, \\
 \Theta_{CCW}(\omega) & = & \frac{k_-}{k_- - i \omega},
\end{eqnarray}
and the power spectrum is 
\begin{equation}
 S_2(\omega) = \frac{2 k_+ k_-}{(k_+ + k_-)} \frac{1}{\omega^2 + (k_+ + k_-)^2}.
\end{equation}
\begin{figure}[t]
  \centering
     \includegraphics[width=0.5\textwidth]{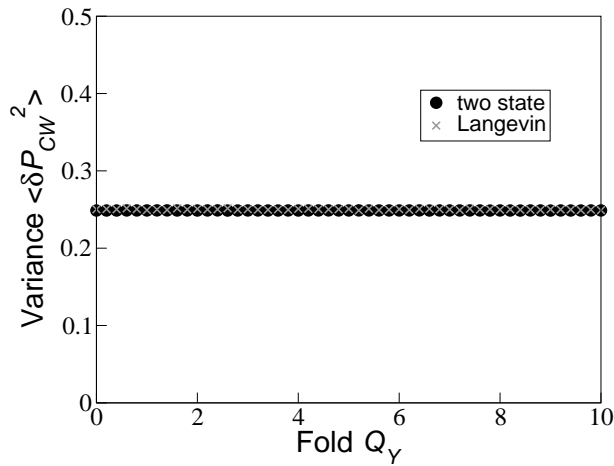}
     \caption{Variance of the motor bias as a function of CheY-P noise intensity $\alpha Q_Y=2\alpha k_Y$ for Langevin and two-state dynamics.\label{fig:varP_Q}}
\end{figure}
This result is equivalent to the spectrum obtained from the Langevin equation.
Furthermore, we tested numerically that the statistics of the Langevin equation and binary process are equivalent for fluctuating rate constants $k_+$ and $k_-$ due to the CheY-P dynamics. We simulated time courses of CheY-P according to the simplified equation
\begin{equation}
 \frac{dN_{Y_p}}{dt}=k_Y-k_{-Y}N_{Y_p}+\eta_{Y_p}(t)
\end{equation}
with rates $k_Y=5/s$ and $k_Y$ such that $\langle N_{Y_p}\rangle^*/V_\text{cell}=\langle Y_p\rangle^*=3.2\mu M$.
The noise term $\eta_{Y_p}(t)$ is Gaussian and white with zero mean and autocorrelation $\langle \eta_{Y_p}(t) \eta_{Y_p}(t') \rangle = 2k_Y\alpha \delta(t-t')\equiv Q_{Y}\alpha \delta(t-t')$, where we varied $\alpha$.
Fluctuating CheY-P was translated into the rates $k_+(Y_p)$ and $k_-(Y_p)$ according to Fig.~2C in the main text. The Langevin equation was simulated using a Euler-Maruyama algorithm~\citep{kloePlat92} and the binary process using a Gillespie algorithm. Figure~\ref{fig:varP_Q} shows the variances of both processes as obtained from $10^2$ runs for each value of $\alpha$. As can be seen from the figure, the Langevin equation is a good description for the binary process of motor switching.

\section{Parameters\label{sec:app:parametersFullModel}}

Rate constants and total cell concentrations of proteins for the full pathway model are given in Table~\ref{tab:fullpathwaymodel}.
The noise terms $\eta_{A_j}$, $\eta_{c_j}$, $\eta_{M_j}$ and $\eta_{P_\text{CW}}$ are the same as in Eq.~(15) and (18)-(20) in the main text and their power spectra are given there.
The noise associated with phosphorylation and dephosphorylation $\eta_{A_p}$, $\eta_{A, B_p}$,$\eta_{A, Y_p}$, $\eta_{-B_p}$ and $\eta_{-Y_p}$ are assumed to be Gaussian white noise terms with zero mean and autocorrelations
$\langle \eta_i(t) \eta_i(t')\rangle = Q_i \delta(t-t')$ with noise intensities $Q_i$ given in Table~\ref{tab:noiseintensities}.
The linearised rate constants for the full pathway model are given in Table~\ref{tab:linearizedrateconstants}.
Fitting parameters of the Fourier transformed response function Fig.~2 are listed in Table~\ref{tab:fitmotorresponsefunction}.
Parameters for Fig.~4 are listed in Tables~\ref{tab:figmotorspec_predictions} and \ref{tab:figmotorspec_korobkova}, and those for Fig.~7 are listed in Table~\ref{tab:FRT}.

\begin{table}[!h]
\caption{\label{tab:fullpathwaymodel}Parameters of the full pathway model, including references to literature.
The literature values are given in parentheses where different from our parameter values. $k_{-Y}$ was determined by the condition that at steady-state with $A_R^*$=1/3, the concentration $[Y_p]^*=[Y]_{\text{tot}}/3$~\citep{SouBerg02}.}
\begin{center}
\begin{tabular}{|l|l|l|}
\hline
Parameter		& Value & Reference\\
\hline
$[A]_{\text{tot}}$ 		& 5 $\mu$M & \citet{SouBerg02}		\\
$[B]_{\text{tot}}$		& 0.28 $\mu$M & \citet{LiHaz04}		\\
$[Y]_{\text{tot}}$ 		& 9.7 $\mu$M& 	\citet{LiHaz04}		\\
$V_{\text{cell}}$		& 1.4 fl & \citet{SouBerg02}		\\
$N_{A,{\text{tot}}}$ 		& 4215 & calculated from above \\
$N_{B,{\text{tot}}}$		& 236 & calculated from above \\
$N_{Y,{\text{tot}}}$ 		& 8177 & calculated from above	\\
$N_\text{tot}=N N_c$		& 7027 & \citet{SouBerg02,LiHaz04}	\\
$k_2$				& 10$^3$ s$^{-1}$ & \citet{ShaLes04}	\\
$k_A$		 		& 10 s$^{-1}$ & \citet{WolBakStock06}		\\
$k_Y$ 				& 100 $\mu$M$^{-1}$ s$^{-1}$  & \citet{StewJahrPar00}\\
$k_B$ 				& 15 $\mu$M$^{-1}$ s$^{-1}$ & \citet{StewJahrPar00}	\\
$k_{-Y}$			& 5 s$^{-1}$	& adjusted to yield steady-state value\\
$k_{-B}$			& 1.35 s$^{-1}$	& (0.35 s$^{-1}$)	\citet{BrayBou95,Stew93}	\\
$\gamma_R$			& 0.006	s$^{-1}$ & \citet{ClauOleEnd10}	\\
$\gamma_B$			& 3.14 $\mu$M$^{-2}$ s$^{-1}$ & \citet{ClauOleEnd10}\\
\hline
\end{tabular}
\end{center}
\end{table}

\begin{table}[!h]
\caption[Intensities of noise terms in full pathway model.]{\label{tab:noiseintensities}Intensities of Gaussian white noise terms in the full pathway model. Index $i$ represents noise term $\eta_i$.}
 \begin{center}
\begin{tabular}{|l|l|l|}
\hline
process & index $i$	&	noise intensity $Q_i$\\ 
\hline
receptor switching & $a$		&	$2 k_2 A^*$\\
ligand diffusion & $L$			&	$2 D s c_0$\\
receptor de/methylation & $M$		&	$2 \gamma_R (N - A^*)$\\
CheA autophosphorylation & ${A_p}$	&	$A_{c}^* \left( \frac{k_A}{N_c N}\right) (N_{A, \text{tot}} - N_{A_p}^*)$\\
CheY phosphorylation & ${A, Y_p}$	&	$\left( \frac{k_y}{V_{\text{cell}}} \right) ( N_{Y, \text{tot}} - N_{Y_p}^* ) N_{A_p}^*$\\
CheB phosphorylation & ${A, B_p}$ 	&	$\left( \frac{k_b}{V_{\text{cell}}} \right) ( N_{B, \text{tot}} - N_{B_p}^* ) N_{A_p}^*$\\
CheY dephosphorylation & ${-Y_p}$ 	&	$k_{-y} N_{Y_p}^*$\\
CheB dephosphorylation & ${-B_p}$ 	&	$k_{-b} N_{B_p}^*$\\
motor switching & ${P_\text{CW}}$ 		&	$\frac{2 k_+^* k_-^*}{k_+^* + k_-^*}$\\
\hline
\end{tabular}
\end{center}
\end{table}

\begin{table}[!h]
\caption[Parameters of the linearised equations for the full pathway.]{\label{tab:linearizedrateconstants}Parameters of the linearised equations for the full pathway.}
\begin{center}
\begin{tabular}{|l | l|}
\hline
$\lambda_i$ & expression\\
\hline
$\lambda_1$ & 		$\gamma_R + \gamma_B {{B_P}^*}^2$\\
$\lambda_2$ & 		$\left( \frac{k_A}{N_c N}\right) ( N_{A, \text{tot}} - N_{A_p}^* )$\\
$\lambda_3$ & 		$A_c^* \left( \frac{k_A}{N_c N}\right) + \left( \frac{k_y}{V_{\text{cell}}} \right) ( N_{Y, \text{tot}} - N_{Y_p}^* ) + \left( \frac{k_B}{V_{\text{cell}}} \right) ( N_{B, \text{tot}} - N_{B_p}^* )$\\
$\lambda_4$ & 		$\left( \frac{k_Y}{V_{\text{cell}}} \right) N_{A_p}^*$\\
$\lambda_5$ & 		$\left( \frac{k_Y}{V_{\text{cell}}} \right) ( N_{Y, \text{tot}} - N_{Y_p}^* )$\\
$\lambda_6$ & 		$\left( \frac{k_Y}{V_{\text{cell}}} \right) N_{A_p}^* + k_{-Y}$\\
$\lambda_7$ & 		$\frac{1}{V_\text{cell}} \left( (1-P^*) \frac{\partial k_+}{\partial Y_p} - P^* \frac{\partial k_-}{\partial Y_p}   \right)$ \\
$\lambda_8$ & 		${k_+}^* + {k_-}^*$\\
$\lambda_9$ & 		$\frac{2 \gamma_B A^* B_p^*}{V_\text{cell}}$\\
$\lambda_{10}$ & 	$\left( \frac{k_B}{V_{\text{cell}}} \right) N_{A_p}^*$\\
$\lambda_{11}$ & 	$\left( \frac{k_B}{V_{\text{cell}}} \right) (N_{B, \text{tot}} - N_{B_p}^*)$\\
$\lambda_{12}$ & 	$\left( \frac{k_B}{V_{\text{cell}}} \right) N_{A_p}^* + k_{-B}$.\\
\hline
\end{tabular}
\end{center}
\end{table}

\begin{table}[!h]
\caption[Fitting parameters for response function of the full pathway model.]{\label{tab:fitmotorresponsefunction}Fitting parameters for response function of the full pathway model for Fig.~2 in the main text. Motor switching rates where not adjusted when fitting to the data by~\citet{ShimTuBerg10} as the high-frequency response was not measured in these experiments.}
\begin{center}
\begin{tabular}{|l|l|l|l|}
\hline
Parameter & \citealp{BloSegBerg82, SegBloBerg86}  & \multicolumn{2}{|l|}{\citealp{ShimTuBerg10}} \\
	& 	[s$^{-1}$]	& 32$^{\circ}$~C [s$^{-1}$] & 22$^{\circ}$~C [s$^{-1}$]\\
\hline
adaptation: & & & \\
$\lambda_1 (\partial A/\partial M)$ 	& 0.178		& 0.018  & 0.0039\\
$\lambda_9$ 				& 0.0263	& 0.0027 & 5.6 $10^{-4}$ \\
\hline
motor switching: & & & \\
$\lambda_7$ 				& 4.4 $10^{-4}$ & --& --\\
$\lambda_8$ 			& 2.111		& --& --\\
\hline
\end{tabular}
\end{center}
\end{table}

\begin{table}[!h]
\caption{\label{tab:figmotorspec_predictions}Parameters for cell-to-cell variation in Fig.~5A in the main text. For additional parameters see Table~\ref{tab:fullpathwaymodel}.}
\begin{center}
\begin{tabular}{|l|l|l|l|l|}
\hline
Parameter & WT1 & red line & green line & blue line\\
&(black line)& & &\\
\hline
$k_+^*$ [s$^{-1}$] & 1.05 & 52.4 & 1.05 & 1.05 \\
$k_-^*$  [s$^{-1}$]& 1.06 & 53.0 & 1.06 &1.06  \\
$\gamma_R$  [s$^{-1}$] & 0.0069 & 0.0069 & 6.9 $10^{-5}$& 0.0069 \\
$\gamma_B$  [$\mu$M$^{-2}$ s$^{-1}$]& 3.14& 3.14& 3.14 $10^{-2}$ &3.14 \\
$N_\text{tot}$ & 7000& 7000& 7000& 70\\
\hline
\end{tabular}
\end{center}
\end{table}

\begin{table}[!h]
\caption{\label{tab:figmotorspec_korobkova}Parameters for cells with low motor bias in Fig.~5B in the main text. For additional parameters see Table~\ref{tab:fullpathwaymodel}. Similar spectra with increased low-frequency component have been found for cells with other motor biases~\citep{ParkPonClu10}.}
\begin{center}
\begin{tabular}{|l|l|}
\hline
Parameter & value\\
\hline
$N_\text{tot}$ & 700 \\
$Y_\text{tot}$ [$\mu$M]& 2 \\
$k_+^*$ [s$^{-1}$] & 0.015\\
$\partial k_+/\partial Y_p$  [s$^{-1}$ $\mu$M$^{-1}$]& 4.75 \\
\hline
\end{tabular}
\end{center}
\end{table}

\begin{table}[!h]
\caption{\label{tab:FRT}Parameters for the fluctuation-response theorem in Fig.~7 in the main text. For additional parameters see Table~\ref{tab:fullpathwaymodel}.}
\begin{center}
\begin{tabular}{|l|l|l|l|}
\hline
Parameter & varying $\gamma_R$ (solid) & varying $\gamma_B$ (dashed) & varying [$Y$]$_\text{tot}$ (dotted) \\
\hline
$N_\text{tot}$ & 700 & 700 & 700 \\
$\gamma_R$ [s$^{-1}$] & varied & 0.0069 & 0.0069 \\
$\gamma_B$  [s$^{-1}$]& 3.14 & varied  & 3.14 \\
$[Y]_{\text{tot}}$ [$\mu$M] & 9.7 & 9.7 & varied\\
\hline
\end{tabular}
\end{center}
\end{table}